\documentclass{article}
\usepackage{authblk} 
\usepackage{graphicx} 
\usepackage{amsfonts}
\usepackage{hyperref}

\usepackage{amsmath, latexsym}
\usepackage{subcaption}
\usepackage{placeins} 
\usepackage{caption}
\usepackage{dsfont}

\title{Exact analytic solutions for the $n=1$ chameleon field in planar systems with an arbitrary number of uniform layers}

\author[1]{C. Briddon \thanks{\texttt{chad.briddon@tuwien.ac.at}}}
\author[1]{M. Pitschmann \thanks{\texttt{mario.pitschmann@tuwien.ac.at}}}

\affil[1]{Atominstitut, Technische Universit\"at Wien, Stadionallee 2, 1020 Vienna, Austria}

\date{}

\begin{document}

\maketitle

\begin{abstract}
    Scalar fields have long been investigated as a potential dark energy candidate. Of these models, screened scalar fields like the chameleon have received interest due to their ability to avoid observational constraints. However, the non-linear nature of these models has led to there being only a small number of exact analytic solutions within the literature. In this work, we present exact analytic solutions for the $n=1$ chameleon model in a planar system consisting of $N$ discrete regions of uniform density. We compare these to the relevant approximate solutions within the literature. We also use our solutions to forecast constraints from \textsc{Cannex}, a recent experiment devised to measure Casimir forces and detect hypothetical fifth forces. Unlike in previous works, we consider the full fifteen-layered setup of \textsc{Cannex}, as opposed to treating the plate and wall as single uniform objects. Doing so revealed an increase in the induced chameleon pressure of up to $\sim 7 \%$. However, the resulting gain is insufficient to probe new regions of parameter space at the projected sensitivities.
\end{abstract}

\section{Introduction}

It has been suggested that the dark energy driving our observed cosmic expansion is not a cosmological constant, but instead a scalar field~\cite{Zlatev:1999PRL, Copeland:2006IJMPD, Joyce:2015PhysRep}.  This idea has received renewed attention due to the results of the DESI collaboration~\cite{abdul2025desi}, which show a preference for a dark energy equation of state that has evolved over the course of cosmic history. Such evolution is not possible for a pure cosmological constant (the most popular proposed source of dark energy), but arises naturally for a dynamical scalar field. One way to independently verify the existence of such a field would be to observe forces mediated by it between masses (called a `fifth force'). However, no such force has been observed, therefore placing strong constraints on the field's coupling strength. This has led to interest in so-called `screened scalar fields', which as the name suggests possess mechanisms that partially screen the masses of objects. In this work we will discuss the chameleon model, in which the field's mass depends on the local matter density~\cite{khoury2004chameleon}. Consequently, in regions close to a dense source, such as the Earth, the field has a larger mass and so any resultant fifth force is suppressed compared to less dense regions. The problem that emerges is that the non-linear equations of motion that give this behaviour are difficult to solve. Even for highly symmetric systems, like spheres and cylinders, only approximate solutions exist~\cite{khoury2004chameleon, burrage2015probing}. One would expect to find exact solutions for systems with planar symmetry, given its simplicity. In fact,~\cite{ivanov2013influence} contains exact solutions for the cases of a single plate located next to a perfect vacuum, and for two parallel plates separated by perfect vacuum. However, the assumption of a perfect vacuum is unrealistic and so these solutions act more as approximate solutions for when the density of the vacuum is negligible. If we wish not to neglect the non-zero vacuum density, there are approximate analytic solutions for a pair of parallel mirrors in~\cite{burrage2016proposed, lemmel2015neutron, brax2010tuning}. An exact solution for a pair of parallel plates with a non-zero vacuum between them was formulated in~\cite{ivanov2016exact}. However, we discovered mistakes in the derivation of this solution. The details of these mistakes are shown in Appendix~\ref{sec : Apx - Correcting previous derivation}, where we also show that the solution in~\cite{ivanov2016exact} can be treated as an approximate solution for cases when the maximum field value is much smaller than its expectation value in the vacuum.

The interest in these solutions comes from their applicability to experiments, thereby allowing us to test for the existence of the chameleon field. \textsc{Cannex}~\cite{sedmik2021next, Haghmoradi:2024ytw}, for example, is designed to measure the induced pressure between a pair of parallel plates caused by Casimir forces, as well as hypothetical fifth forces. Solving for the chameleon field for such a system allows us to compute the corresponding induced pressure, and hence either discover the chameleon or constrain its parameters. Another experiment, \textit{q}BOUNCE~\cite{ivanov2013influence, jenke2014gravity}, has neutrons bounce above a flat horizontal mirror. Because the neutron is a quantum particle, the allowed energy levels are discrete. Measuring the transition rates between energy levels therefore provides information about the local gravitational field. The existence of a scalar field like the chameleon would change these energy levels, thereby changing the transition rates, as was studied in~\cite{ivanov2013influence, jenke2014gravity, brax2011strongly, brax2013probing}. Finally, although not truly planar symmetric, torsion-balance experiments such as E$\rm\ddot{o}$t-Wash have made use of planar solutions as an approximation to produce constraints on the chameleon~\cite{upadhye2012dark}.

Ultimately all the existing solutions for planar systems invoke some approximation, whether that be neglecting the non-zero vacuum density, or taking some parameter like the Compton wavelength in certain regions to be small. Naturally, actual experimental setups are much more complicated than these solutions allow, with a single plate usually being comprised of multiple layers of different materials. We would expect this caveat to be negligible when the field's Compton wavelength within the plate is either large enough so that the field interacts with an effective density that is an average over the plate layers, or small enough so the plate is screened and only the density of the layer next to the vacuum effectively contributes. For intermediate wavelengths, however, the actual structure of the plates will affect the overall field profile. In this work we present an exact solution for the ($n=1$) chameleon field in a planar symmetric system consisting of $N$ discrete layers of uniform density (the width and densities of each layer are arbitrary). Specific attention will be paid to the $N = 2, 3, 4$ cases, linking them to previous results in the literature. Finally, our solution will be applied to \textsc{Cannex}, taking into account every sub-layer that composes the corresponding plates. We then use this analytic solution to forecast the constraints that \textsc{Cannex} will place on the chameleon, once it is operational, and show how the constrained region varies with the size of the vacuum gap and gas density. 

Throughout this work we will be working in natural units ($c=\hbar=1$), and will use the mostly positive metric signature, $\eta_{\mu \nu} = {\rm diag} (-1, 1, 1, 1)$. The Planck mass, in natural units, has the value $M_{\rm pl} := (8\pi G_N)^{-1/2} \approx 2.4 \cdot 10^{18} \, \rm{GeV}$, where $G_N$ is Newton's constant.

\section{The chameleon model}
\label{sec : The chameleon model}

The equation of motion of the chameleon field is
\begin{equation}
    \label{eq : chameleon e.o.m. (general)}
    \Box \phi = \frac{\partial V_{\rm eff}(\phi)}{\partial \phi}, \qquad V_{\rm eff}(\phi) = \Lambda^4 \left( 1 + \frac{\Lambda^{n}}{\phi^n} \right) + \frac{\rho}{M} \phi,
\end{equation}
where $n$ is an integer, while $\Lambda$ and $M$ are model parameters to be determined/constrained through experiments~\cite{khoury2004chameleon, burrage2018tests}. To give physical meaning to the model parameters, $M$ is related to the coupling strength between the chameleon field and matter $\beta$ by the relation $M := M_{\rm pl}/\beta$, while $\Lambda$ is taken to be the apparent dark energy contribution of the field. In general $n > 0$ or $n = -4, -6, -8, ...$~\cite{burrage2018tests}; however, we shall work exclusively with the $n=1$ case. Additionally, we will only concern ourselves with systems which are static, and invariant under transformations in the $x-y$ plane, i.e. which are 1-dimensional. Therefore, the equation we desire to solve is in fact
\begin{equation}
    \label{eq : chameleon e.o.m.}
    \phi''(z) = -\frac{\Lambda^5}{\phi^2(z)} + \frac{\rho(z)}{M},
\end{equation}
where $\phi'$ and $\phi''$ will hence be used to denote the first and second derivatives of the field with respect to the coordinate $z$.

In each region the field will move towards the minimum of its effective potential. This value is referred to as the expectation value (EV) and is given by
\begin{equation}
    \label{eq : expectation Value (EV)}
    \phi^{\rm EV} := \sqrt{\frac{M \Lambda^5}{\rho}}.
\end{equation}
With this we can compute the effective mass of the field around its potential minimum as 
\begin{equation}
    \label{eq : Chameleon mass}
    m_\phi^2 := V_{\rm eff}''(\phi^{\rm EV}) = \frac{2 \Lambda^5}{\left( \phi^{\rm EV} \right)^3} = \frac{2}{\Lambda^{5/2}} \left( \frac{\rho}{M} \right)^{3/2}.
\end{equation}
The important detail in this expression is that the chameleon's mass has an explicit $\rho$ dependence: this is the defining property of the chameleon mechanism. In regions with greater density $\rho$ the effective mass of the field is larger, and so forces mediated by it are suppressed relative to regions with lower density. An alternative way to view it is to recall that the field's Compton wavelength (the length scales over which the field will vary) is the inverse of the field mass, and is therefore
\begin{equation}
    \label{eq : chameleon Compton wavelength}
    \lambda = \sqrt{\frac{\left( \phi^{\rm EV} \right)^3}{2 \Lambda^5}} = \frac{\Lambda^{5/4}}{\sqrt{2}} \left( \frac{M}{\rho} \right)^{3/4}.
\end{equation}
Consequently, as $\rho$ increases $\lambda$ decreases and so does the effective range of the chameleon fifth force, which for a test particle of mass $m_p$ is
\begin{equation}
    \label{eq : chameleon fifth force}
    \vec{F}_c = - \frac{m_p}{M} \vec{\nabla} \phi.
\end{equation}

\section{Solution for $N$-layer system}
\label{sec : Solution for N-layer system}

The class of systems we are solving Eq.~(\ref{eq : chameleon e.o.m.}) for are invariant in the $x-y$ plane and consist of $N$ regions of uniform density separated in the z-direction. The length and density of each region are $L_i$ and $\rho_i$ respectively, where the region label $i \in \{1, 2, ..., N\}$ runs from left to right (increasing in $z$). Note that the exterior regions, $i=1$ and $i=N$, will extend infinitely in the negative and positive $z$-directions respectively. Since we know $\rho_i$ we also know the EV in each region, which we shall label as $\phi^{\rm EV}_i$.

The behaviour of the field in the exterior and interior regions will differ from one another. We will therefore address these two scenarios separately before unifying them using the continuity of the field at their interfaces.

\subsection{Exterior regions}
\label{sec : Solution for N-layer system - exterior regions}

Since the exterior regions have infinite extension, we can use the boundary conditions $\phi \rightarrow \phi^{\rm EV}_i$ and $\phi' \rightarrow 0$ as $|z| \rightarrow \infty$, for $i\in\{1,N\}$. By integrating Eq.~(\ref{eq : chameleon e.o.m.}) with these boundary conditions, we find
\begin{equation}
    \label{eq : Exterior regions - first integration}
    \frac{1}{2} \left( \phi' \right)^2
    = V_{\rm eff}(\phi) - V_{\rm eff}(\phi^{\rm EV}_i) = \frac{\Lambda^5}{\left( \phi^{\rm EV}_i \right)^2} \frac{\left( \phi - \phi^{\rm EV}_i \right)^2}{\phi}.
\end{equation}
With some rearranging we obtain
\begin{equation}
    \label{eq : Exterior regions - field gradient}
    \phi' = \left( \delta_{1i} - \delta_{Ni} \right) \, \frac{\sqrt{\phi^{\rm EV}_i}}{\lambda_i} \frac{ \left( \phi - \phi^{\rm EV}_i \right)}{\sqrt{\phi}},
\end{equation}
where $\lambda_i$ is simply Eq.~(\ref{eq : chameleon Compton wavelength}) evaluated at $\rho=\rho_i$. Note the choice of sign, $\left( \delta_{1i} - \delta_{Ni} \right)$ where $\delta_{ij}$ is the Kronecker delta, comes from the fact that the field approaches $\phi^{\rm EV}_i$ as $|z| \rightarrow \infty$ in both exterior regions. We now wish to integrate Eq.~(\ref{eq : Exterior regions - field gradient}) between the boundary at the interface, located at $z=z_b$ and with field value $\phi_b$, and some arbitrary $z$, which gives
\begin{equation}
    \label{eq : Exterior regions - second integration}
    \begin{aligned}
        -\frac{|z - z_b|}{\lambda_i} &= \frac{1}{\sqrt{\phi^{\rm EV}_i}} \int_{\phi_b}^{\phi} \frac{\sqrt{\varphi}}{\varphi - \phi^{\rm EV}_i} d\varphi \\
        &= 2 \int_{u_b}^{u} \frac{t^2}{t^2 - 1} dt \\
        &= 2 (u-u_b) + \ln\left( \frac{u-1}{u_b-1} \right) - \ln\left( \frac{u+1}{u_b+1} \right),
    \end{aligned}
\end{equation}
where on the second line we have used the substitution $u:=\sqrt{\phi/\phi^{\rm EV}_i}$. Therefore, the field profile inside the exterior regions must obey the equation
\begin{equation}
    \label{eq : Exterior regions - field profile}
    \sqrt{\frac{\phi}{\phi^{\rm EV}_i}} - \sqrt{\frac{\phi_b}{\phi^{\rm EV}_i}} + \frac{1}{2} \ln \left( \frac{\sqrt{\phi} - \sqrt{\phi^{\rm EV}_i}}{\sqrt{\phi_b} - \sqrt{\phi^{\rm EV}_i}} \, \frac{\sqrt{\phi_b} + \sqrt{\phi^{\rm EV}_i}}{\sqrt{\phi} + \sqrt{\phi^{\rm EV}_i}} \right) = - \frac{1}{2 \lambda_i} |z - z_b|.
\end{equation}
Unfortunately this equation cannot be inverted in closed form, meaning we must leave it in the form $z=z(\phi)$. That said, since the field is monotonic, which is evident from Eq.~(\ref{eq : Exterior regions - field gradient}), and bounded between $\phi_b$ and $\phi^{\rm EV}_i$, the map from $\phi$ to $z$ is one-to-one. Therefore, there is no issue using Eq.~(\ref{eq : Exterior regions - field profile}) in its current form. We also note that the field value at the boundary, $\phi_b$, is the only unknown in Eq.~(\ref{eq : Exterior regions - field profile}), and therefore determines the shape of the exterior profile.

\subsection{Interior regions}
\label{sec : Solution for N-layer system - Interior regions}

Unlike in the exterior regions, here we do not have a predetermined boundary condition. Instead we will take the field at the left and right boundaries to be $\phi_L$ and $\phi_R$ respectively, and will later determine them using the continuity of the field at said boundaries. Since we are solving a second order equation, the two boundary values are enough information to uniquely determine the field profile. The allowed solutions then fall into one of three possibilities, depending on the hierarchy of $\phi_L$, $\phi_R$, and $\phi^{\rm EV}_i$. In the following sections we will solve Eq.~(\ref{eq : chameleon e.o.m.}) for each of these scenarios. Note that in each case we will begin with a z-coordinate whose origin lies at the centre of the region in question. This is always valid because we know the positions of the region boundaries a priori.

\subsubsection{Solution with a maximum}
\label{sec : Solution for N-layer system - Interior regions - Solution with a maximum}

Let us first consider solutions which possess a maximum of $\phi_0$. To make the calculation easier we will construct a new coordinate $\zeta$, which by definition has the location of $\phi_{0}$ at the origin. The position of the left and right boundaries are labelled as $-\zeta_L$ and $+\zeta_R$, respectively. The transformation between the two sets of coordinates is then
\begin{equation}
    \label{eq : interior region - coorinate transformation (z to zeta)}
    \zeta := z + \frac{(\zeta_R - \zeta_L)}{2}.
\end{equation}
Note that $\zeta_L$ and $\zeta_R$ will depend on the model parameters but their sum is $L_i = \zeta_L + \zeta_R$; recall $L_i$ is the physical separation between the boundaries. If the field profile has its maximum inside the interior region then both $\zeta_R$ and $\zeta_L$ will be positive.

Integrating Eq.~(\ref{eq : chameleon e.o.m.}) between $\zeta=0$ and some arbitrary $\zeta$ gives
\begin{equation}
    \label{eq : chameleon e.o.m. - integrated once (extrema)}
    \frac{1}{2} \left( \frac{\partial \phi}{\partial \zeta} \right)^2
    = V_{\rm eff}(\phi) - V_{\rm eff}(\phi_0) = \frac{\Lambda^5}{\phi_0} \frac{(1 - x)(1 - k^2 x)}{x},
\end{equation}
where $x := \phi/\phi_0$ and $k := \phi_0/\phi^{\rm EV}_i$. We can then express the field gradient as
\begin{equation}
    \label{eq : interior region - field gradient (maximum)}
    x'(\zeta) = -\frac{{\rm sgn}(\zeta)}{\lambda_i k^{3/2}} \sqrt{\frac{(1 - x) (1 - k^2 x)}{x}},
\end{equation}
where our choice of sign encodes that the maximum is at $\zeta=0$ (when $x=1$). Since the gradient must be real-valued, we have $(1 - x) (1 - k^2 x) \geq 0$. Considering a small displacement from the maximum, $x = 1 - \delta$, the inequality becomes $\delta \cdot (1 - k^2 + k^2 \delta) \geq 0$. If this inequality is to hold for arbitrarily small values of $\delta$ then we must have that $k^2 \leq 1$. But in Section~\ref{sec : Solution for N-layer system - exterior regions} we solved the $k=1$ case and found the solution approached $\phi^{\rm EV}_i$ asymptotically. If we require the position of the maximum to be a finite value, then we have $k^2 < 1$ and therefore $\phi_0 < \phi^{\rm EV}_i$. In other words, the field profile will only permit a solution with a maximum if $\max(\phi_L, \phi_R) \leq \phi_0 < \phi^{\rm EV}_i$. Note that this excludes the possibility of a second extremum at $x = 1/k^2$, meaning the field can either possess a single maximum or be monotonic. If the solution is monotonic, the maximum used to define our $\zeta$-coordinate is now treated as a fictitious maximum that lies outside the region in question. Consequently, this would make either $\zeta_L$ or $\zeta_R$ negative.

Integrating Eq.~(\ref{eq : interior region - field gradient (maximum)}) between the maximum and some arbitrary point gives
\begin{equation}
    \label{eq : interior region - field gradient (maximum) - integrated}
    \begin{aligned}
        \frac{|\zeta|}{2 \lambda_i k^{3/2}}
        &= \frac{1}{2} \int_{x}^{1} \sqrt{\frac{t}{(1-t) (1 - k^2 t)}} \, dt \\
        &= \int_{{\rm sn}^{-1} \left( \sqrt{x}, k \right)}^{{\rm K} \left( k \right)} \frac{{\rm sn}^2(u, k) \, {\rm cn}(u, k) \, {\rm dn}(u, k)}{\sqrt{(1-{\rm sn}^2(u, k)) (1 - k^2 {\rm sn}^2(u, k))}} \, du \\
        &= \int_{{\rm sn}^{-1} \left( \sqrt{x}, k \right)}^{{\rm K} \left( k \right)} {\rm sn}^2(u, k) \, du \\
        &= {\rm Sn}\left(K(k), k \right) - {\rm Sn}({\rm sn}^{-1}\left(\sqrt{x}, k), k\right),
    \end{aligned}
\end{equation}
where we have invoked the Jacobi elliptic functions defined in Appendix~\ref{sec : Apx - Jacobi Elliptic Functions}. Specifically, we used the substitution $t = {\rm sn}^2(u, k)$ on the second line (see Eqs.~(\ref{eq : Apx - Jacobi elliptic functions : Elliptic integral - complete first kind}) and (\ref{eq : Apx - Jacobi elliptic functions : sn (definition)}) for the definitions of $K$ and ${\rm sn}$, respectively), on the third line we used the identities seen in Eqs.~(\ref{eq : Apx - Jacobi elliptic functions : dn (definition)}) and (\ref{eq : Apx - Jacobi elliptic functions : identity - sn to cn}), and in the final line we used the definition of ${\rm Sn}$ as defined in Eq.~(\ref{eq : Apx - Jacobi elliptic functions : integral squared naming rule}). Therefore, the expression for the field profile is
\begin{equation}
    \label{eq : interior region - maximum - field}
    \phi = \phi_0 \, {\rm sn}^2 \left( {\rm Sn}^{-1} \left\{ \left[ {\rm Sn} \left( K(k), k \right) - \frac{1}{2 \lambda_i k^{3/2}} \left| z + \frac{(\zeta_R - \zeta_L)}{2} \right| \right], k \right\}, k \right).
\end{equation}
Our final step to completing this solution is to determine $\zeta_L$, $\zeta_R$, and $k$, from our boundary values $\phi_L$ and $\phi_R$. By evaluating Eq.~(\ref{eq : interior region - maximum - field}) at the boundaries, we get the relation
\begin{equation}
    \label{eq : interior region - maximum - zeta (magnitude)}
    |\zeta_j| = 2 \lambda_i k^{3/2} \left[ {\rm Sn} \left\{ K(k), k \right\} - {\rm Sn} \left\{ {\rm sn}^{-1} \left( \sqrt{x_j}, k \right), k \right\} \right],
\end{equation}
where $x_j := \phi_j/\phi_0$ and $j \in \{L, R\}$. Note that Eq.~(\ref{eq : interior region - maximum - zeta (magnitude)}) only gives the magnitude of $\zeta_j$, with its sign needing to be ascertained separately. As mentioned previously, when both $\zeta_L$ and $\zeta_R$ are positive then the profile has a maximum, while a monotonic profile means either $\zeta_L$ or $\zeta_R$ is negative. Using the constraint $L_i = \zeta_L + \zeta_R$, if $\zeta_L$ is negative then $\zeta_R > L_i$ and vice versa. If we apply the latter inequality to Eq.~(\ref{eq : interior region - maximum - zeta (magnitude)}) we find that if $|\zeta_j| > L_i$ then $\sqrt{x_c} > \sqrt{x_j}$, where
\begin{equation}
    \label{eq : interior region - maximum - monotonic constraint}
    \sqrt{x_c} = {\rm sn} \left( {\rm Sn}^{-1} \left\{ \left[ {\rm Sn} \left( K(k), k \right) - \frac{L_i}{2 \lambda_i k^{3/2}} \right], k \right\}, k \right).
\end{equation}
Therefore, we can express the values $\zeta_L$ and $\zeta_R$ as
\begin{equation}
    \label{eq : interior region - maximum - zeta (sign)}
    \begin{aligned}
        & \zeta_L = |\zeta_L|, \qquad \zeta_R = {\rm sgn} \left( \sqrt{x_L} - \sqrt{x_c} \right) |\zeta_R|, \qquad \text{if } x_L < x_R \\
        & \zeta_R = |\zeta_R|, \qquad \zeta_L = {\rm sgn} \left( \sqrt{x_R} - \sqrt{x_c} \right) |\zeta_L|, \qquad \text{if } x_L > x_R.
    \end{aligned}
\end{equation}
Applying this to our constraint, $L_i = \zeta_L + \zeta_R$, we have
\begin{equation}
    \label{eq : interior region - maximum - constraint equation (k)}
    \begin{aligned}
         f_{1}(x_L, x_R, k) := k^{3/2} \bigg| & (1 + \sigma) \, {\rm Sn} \left\{ K(k), k \right\} - \sigma \, {\rm Sn} \left\{ {\rm sn}^{-1} \left( \sqrt{x_R}, k \right), k \right\} \\
        & - {\rm Sn} \left\{ {\rm sn}^{-1}\left( \sqrt{x_L}, k \right), k \right\} \bigg| = \frac{L_i}{2 \lambda_i},
    \end{aligned}
\end{equation}
where $\sigma := {\rm sgn}\left({\rm min}\left(\sqrt{x_L}, \sqrt{x_R} \right) - \sqrt{x_c} \right)$. Therefore, if we fix the values of $x_L$ and $x_R$, we can solve Eq.~(\ref{eq : interior region - maximum - constraint equation (k)}) for the corresponding value of $k$. From there, we can compute $\zeta_L$ and $\zeta_R$ using Eqs.~(\ref{eq : interior region - maximum - zeta (magnitude)}) and (\ref{eq : interior region - maximum - zeta (sign)}), which means we can determine the field profile in Eq.~(\ref{eq : interior region - maximum - field}) entirely from knowing the field values at the boundaries.

Above, we treated $\phi_L$ and $\phi_R$ as independent inputs to the field profile. However, the rate at which the field can vary is limited by its Compton wavelength. Consequently, some combinations of $\phi_L$ and $\phi_R$ are not accessible to our solution in Eq.~(\ref{eq : interior region - maximum - field}). To show this, we can compare the field gradient in Eq.~(\ref{eq : interior region - field gradient (maximum)}) for some $k < 1$ against the $k \rightarrow 1$ limit, which is the exterior solution given by Eq.~(\ref{eq : Exterior regions - field gradient}). Doing so, we find that when we evaluate both gradients at the same field value, the exterior solution always possesses the larger gradient magnitude. Therefore, the exterior solution represents the limit for the shortest distance between any two field values in this scenario. This minimum length can be derived by evaluating Eq.~(\ref{eq : Exterior regions - field profile}) using both boundary values, which gives the expression
\begin{equation}
    \label{eq : interior region - maximum - allowed scenario (L^star)}
    L^*_i(\phi_L, \phi_R) = 2 \lambda_i \left| \sqrt{\frac{\phi_L}{\phi^{\rm EV}_i}} - \sqrt{\frac{\phi_R}{\phi^{\rm EV}_i}} + \frac{1}{2} \ln\left( \frac{\sqrt{\phi^{\rm EV}_i} - \sqrt{\phi_L}}{\sqrt{\phi^{\rm EV}_i} - \sqrt{\phi_R}} \right) + \frac{1}{2} \ln\left( \frac{\sqrt{\phi^{\rm EV}_i} + \sqrt{\phi_R}}{\sqrt{\phi^{\rm EV}_i} + \sqrt{\phi_L}} \right) \right|.
\end{equation}
Therefore, if $L_i > L^*_i$ is satisfied then $\phi_L$ and $\phi_R$ are valid boundary values, and our field profile is given by Eq.~(\ref{eq : interior region - maximum - field})\footnote{Note that Eq.~(\ref{eq : interior region - maximum - allowed scenario (L^star)}) is not defined when $\phi^{\rm EV}_i$ lies between $\phi_L$ and $\phi_R$ (inclusive). In this case the field must cross
the EV, which rules out the extremum solution, since solutions possessing an extremum never cross the EV. We therefore set $L^*_i = \infty$ in this case, which ensures that $L_i > L^*_i$ is never satisfied and is consistent with $L^*_i$ diverging as $\phi_L$ or $\phi_R$ approaches $\phi^{\rm EV}_i$.}. If this condition is not satisfied (the layer in question is too small) then a new solution is needed. Specifically, the curve would need to be steeper than the exterior solution, which is only possible if it crosses the field EV (inside or outside the region). These solutions are discussed in Section~\ref{sec : Solution for N-layer system - Interior regions - Profile that passes EV}.

\subsubsection{Solution with a minimum}
\label{sec : Solution for N-layer system - Interior regions - Solution with a minimum}

Let us now consider the possibility that the field profile possesses a minimum. We can use the same coordinate system we used in Section~\ref{sec : Solution for N-layer system - Interior regions - Solution with a maximum}, except now the origin ($\zeta = 0$) is the location of the minimum. We will use $\phi_0$ as the value of the minimum and continue to use the same definitions for $x$ and $k$ as before. The derivative equation we wish to solve is therefore
\begin{equation}
    \label{eq : interior region - field gradient (minimum)}
    x'(\zeta) = +\frac{{\rm sgn}(\zeta)}{\lambda_i k^{3/2}} \sqrt{\frac{(1 - x) (1 - k^2 x)}{x}}.
\end{equation}
Aside from the sign change from Eq.~(\ref{eq : interior region - field gradient (maximum)}), the main difference is that $x \geq 1$, with again $x=1$ at $\zeta=0$. As before, we require $(1-x)(1 - k^2x) \geq 0$. If we consider the perturbation $x = 1 + \delta$ we then have $\delta (1 - k^2 - k^2 \delta) \leq 0$. If this is to hold for arbitrarily small values of $\delta$ then we must have $k^2 > 1$ (since $k \neq 1$ as mentioned previously), and therefore $\phi_0 > \phi^{\rm EV}_i$. The minimum solution is then only valid if $\phi^{\rm EV}_i < \phi_0 \leq \min(\phi_L, \phi_R)$. As before, this excludes a possible second extremum at $x = 1/k^2$ meaning the field either has a single minimum or is monotonic, where in the latter case either $\zeta_L$ or $\zeta_R$ is negative and the minimum lies outside the region in question. 

Motivated by the form of our previous solution, let us integrate Eq.~(\ref{eq : interior region - field gradient (minimum)}) using the substitution $x={\rm ns}^2(u, m)$, where ${\rm ns}$ is defined in Eq.~(\ref{eq : Apx - Jacobi elliptic functions : multiplicative inverse}). Using Eqs.~(\ref{eq : Apx - Jacobi elliptic functions : sn derivative}) and (\ref{eq : Apx - Jacobi elliptic functions : multiplicative inverse}) we can derive that
\begin{equation}
    \label{eq : Derivative of ns^2}
    \frac{d}{du} {\rm ns}^2(u, m) = -2 \, {\rm ns}^3(u, m) \, {\rm cn}(u, m) \, {\rm dn}(u, m),
\end{equation}
and therefore the integral of Eq.~(\ref{eq : interior region - field gradient (minimum)}) is
\begin{equation}
    \label{eq : interior region - field gradient (minimum) - integrated}
        \frac{|\zeta|}{2 \lambda_i k^{3/2}}
        = \frac{1}{k} \int_{{\rm ns}^{-1}\left(\sqrt{x}, m\right)}^{K\left(m\right)} {\rm ns}^2(u, m) \left( \frac{{\rm dn}(u, m)}{\sqrt{1 - k^{-2} {\rm sn}^2(u, m)}} \right) du.
\end{equation}
If we set $m=k^{-1}$, then the bracketed term in the integral is equal to unity due to the identity in Eq.~(\ref{eq : Apx - Jacobi elliptic functions : dn (definition)}). The integral on the right-hand side of Eq.~(\ref{eq : interior region - field gradient (minimum) - integrated}) is then
\begin{equation}
    \label{eq : integral of ns^2}
    \int_{{\rm ns}^{-1}\left(\sqrt{x}, k^{-1}\right)}^{K\left(k^{-1}\right)} {\rm ns}^2(u, k^{-1}) du = {\rm Ns} \left\{ K\left(k^{-1}\right), k^{-1} \right\} - {\rm Ns} \left\{ {\rm ns}^{-1} \left( \sqrt{x}, k^{-1} \right), k^{-1} \right\},
\end{equation}
where ${\rm Ns}$ is defined in Eq.~(\ref{eq : Apx - Jacobi elliptic functions : integral squared naming rule}). Therefore, the field profile which possesses a minimum is
\begin{equation}
    \label{eq : interior region - minimum - field}
    \phi = \phi_0 \, {\rm ns}^2 \left( {\rm Ns}^{-1} \left\{ \left[ {\rm Ns} \left( K \left( \frac{1}{k} \right), \frac{1}{k} \right) - \frac{1}{2 \lambda_i \sqrt{k}} \left| z + \frac{(\zeta_R - \zeta_L)}{2} \right| \right], \frac{1}{k} \right\}, \frac{1}{k} \right).
\end{equation}
To obtain expressions for $\zeta_L$, $\zeta_R$, and $k$ in terms of the boundary values $x_L$ and $x_R$ we will use the same procedure as used in Section~\ref{sec : Solution for N-layer system - Interior regions - Solution with a maximum}. Evaluating Eq.~(\ref{eq : interior region - minimum - field}) at the boundaries gives
\begin{equation}
    \label{eq : interior region - minimum - zeta (magnitude)}
    |\zeta_j| = 2 \lambda_i \sqrt{k} \left[ {\rm Ns} \left\{ K \left( \frac{1}{k} \right), \frac{1}{k} \right\} - {\rm Ns} \left\{ {\rm ns}^{-1} \left( \sqrt{x_j}, \frac{1}{k} \right), \frac{1}{k} \right\} \right].
\end{equation}
Using the condition $|\zeta_j| > L_i$ we have $\sqrt{x_j} > \sqrt{x_c}$ where now
\begin{equation}
    \label{eq : interior region - minimum - monotonic constraint}
    \sqrt{x_c} = {\rm ns} \left( {\rm Ns}^{-1} \left\{ \left[ {\rm Ns} \left\{ K \left( \frac{1}{k} \right), \frac{1}{k} \right\} - \frac{L_i}{2 \lambda_i \sqrt{k}} \right], \frac{1}{k} \right\}, \frac{1}{k} \right),
\end{equation}
and the signs of $\zeta_L$ and $\zeta_R$ are
\begin{equation}
    \label{eq : interior region - minimum - zeta (sign)}
    \begin{aligned}
        & \zeta_L = |\zeta_L|, \qquad \zeta_R = {\rm sgn} \left( \sqrt{x_c} - \sqrt{x_L} \right) |\zeta_R|, \qquad \text{if } x_L > x_R \\
        & \zeta_R = |\zeta_R|, \qquad \zeta_L = {\rm sgn} \left( \sqrt{x_c} - \sqrt{x_R} \right) |\zeta_L|, \qquad \text{if } x_L < x_R.
    \end{aligned}
\end{equation}
Then $L_i = \zeta_L + \zeta_R$ becomes
\begin{equation}
    \label{eq : interior region - minimum - constraint equation (k)}
    \begin{aligned}
         f_{2}(x_L, x_R, k) := \sqrt{k} \, \bigg| & (1 + \sigma) \, {\rm Ns} \left\{ K \left( \frac{1}{k} \right), \frac{1}{k} \right\} - \sigma \, {\rm Ns} \left\{ {\rm ns}^{-1} \left( \sqrt{x_R}, \frac{1}{k} \right), \frac{1}{k} \right\} \\
         & - {\rm Ns} \left\{ {\rm ns}^{-1}\left( \sqrt{x_L}, \frac{1}{k} \right), \frac{1}{k} \right\} \bigg| = \frac{L_i}{2 \lambda_i},
    \end{aligned}
\end{equation}
where $\sigma := {\rm sgn}\left(\sqrt{x_c} - {\rm max}\left(\sqrt{x_L}, \sqrt{x_R}\right) \right)$. We once again can derive the parameters of the field profile completely from the boundary values $x_L$ and $x_R$.

As was the case in Section~\ref{sec : Solution for N-layer system - Interior regions - Solution with a maximum}, not all combinations of $\phi_L$ and $\phi_R$ are accessible to the solution in Eq.~(\ref{eq : interior region - minimum - field}). Following the same logic used in Section~\ref{sec : Solution for N-layer system - Interior regions - Solution with a maximum} once again reveals that the condition for $\phi_L$ and $\phi_R$ to be valid boundary values is $L_i > L^*_i$, where $L^*_i$ is defined in Eq.~(\ref{eq : interior region - maximum - allowed scenario (L^star)}). If this condition is violated, this requires the curve to cross the field EV. We will now discuss these solutions in the section below.

\subsubsection{Profile that passes through the EV}
\label{sec : Solution for N-layer system - Interior regions - Profile that passes EV}

In the previous sub-sections, we found that solutions for the interior regions that possess an extremum (fictitious or otherwise) must satisfy $L_i > L^*_i(\phi_L, \phi_R)$, otherwise the curve must cross the field EV either inside or outside the region in question. In this section we will consider such a scenario. We begin by defining a new coordinate
\begin{equation}
    \label{eq : interior region - coorinate transformation (z to eta)}
    \eta := z - z^*,
\end{equation}
where $z^*$ is the location of $\phi = \phi^{\rm EV}_i$. Now integrating Eq.~(\ref{eq : chameleon e.o.m.}) between our newly defined origin and some arbitrary point gives
\begin{equation}
    \label{eq : interior region - passes EV - first integration (non-zero grad at origin)}
    \frac{1}{2} \left( \frac{d \phi}{d \eta} \right)^2 - \frac{1}{2} \left( \phi_0' \right)^2
    = V_{\rm eff}(\phi) - V_{\rm eff}\left( \phi^{\rm EV}_i \right) = \frac{\Lambda^5}{\left( \phi^{\rm EV}_i \right)^2} \frac{\left( \phi - \phi^{\rm EV}_i \right)^2}{\phi},
\end{equation}
where $\phi_0'$ is the gradient at the origin, which needs to be determined. Notice, if $\phi_0' = 0$ then the above formula reduces to Eq.~(\ref{eq : Exterior regions - first integration}). We know from Section~\ref{sec : Solution for N-layer system - exterior regions} that this leads to a solution that asymptotically approaches $\phi^{\rm EV}_i$, and since this solution is known we will neglect it and only consider when $\phi_0' \neq 0$.

By rearranging Eq.~(\ref{eq : interior region - passes EV - first integration (non-zero grad at origin)}) we find that the field gradient is
\begin{equation}
    \label{eq : interior region - passes EV - field gradient}
    y'(\eta) = {\rm sgn}(y_R - y_L) \frac{2}{\lambda_i} \sqrt{C^2 + \frac{(y-1)^2}{4y}},
\end{equation}
where $y:=\phi/\phi^{\rm EV}_i$, and $C := \lambda_i \phi_0' / 2 \phi^{\rm EV}_i$ is a constant. Our knowledge of the gradient's sign comes from the fact that in this scenario the field must be monotonic since the quantity inside the square root can never be zero (because $\phi_0' \neq 0$ therefore $C^2 > 0$). Integrating this from the origin we find
\begin{equation}
    \label{eq : interior region - passes EV - integral (1)}
    \begin{aligned}
        {\rm sgn}(y_R - y_L) \, \frac{2\eta}{\lambda_i}
        &= \int_{1}^{y} \left[ C^2 + \frac{(t-1)^2}{4t} \right]^{-1/2} dt \\
        &= \int_{1}^{y} \left[ C^2 + \left( \frac{t^{1/2} - t^{-1/2}}{2} \right)^2 \right]^{-1/2} dt \\
        &= \frac{1}{|C|} \int_{1}^{y} \left[ 1 + |C|^{-2} \sinh^2 \left( g(t) \right) \right]^{-1/2} dt,
    \end{aligned}
\end{equation}
where on the third line we used the shorthand $g(t) := \ln(\sqrt{t})$. Next we shall show how the integrand can be expressed in terms of the Jacobi elliptic functions. Beginning with the Jacobi elliptic function ${\rm dc}$, as defined in Eq.~(\ref{eq : Apx - Jacobi elliptic functions : ratio naming rule}), we can show that
\begin{equation}
    \label{eq : interior region - passes EV - dc identity}
    \begin{aligned}
        {\rm dc} \left( u, \sqrt{1-m^2} \right) \equiv {\rm dn}(iu, m) 
        &= \sqrt{1 - m^2 \sin^2 \left( F^{-1}(iu, m) \right)} \\
        &= \sqrt{1 - m^2 \sin^2 \left( i g \right)} \\
        &= \sqrt{1 + m^2 \sinh^2 \left( g \right)},
    \end{aligned}
\end{equation}
where in the first line we used the identities in Eqs.~(\ref{eq : Apx - Jacobi elliptic functions : sn (definition)}), (\ref{eq : Apx - Jacobi elliptic functions : dn (definition)}), and (\ref{eq : Apx - Jacobi elliptic functions : argument u is imaginary (dn)}), in the second line we invoke the identity in Eq.~(\ref{eq : Apx - Jacobi elliptic functions : argument u is imaginary (F)}) by using the substitution $u=F\left({\rm tan}^{-1} \left( {\rm sinh}(g) \right), \sqrt{1 - m^2}\right)$, and on the last line we used $\sinh(x) \equiv -i \sin(ix)$. Comparing Eqs.~(\ref{eq : interior region - passes EV - integral (1)}) and (\ref{eq : interior region - passes EV - dc identity}) we have
\begin{equation}
    \label{eq : interior region - passes EV - integral (2)}
    {\rm sgn}(y_R - y_L) \, \frac{2\eta}{\lambda_i} = \frac{1}{|C|} \int_{1}^{y} {\rm cd} \left( u(t), \mu \right) dt,
\end{equation}
where $\mu := \sqrt{1 - C^{-2}}$, and $u$ is related to $y$ by
\begin{equation}
    \label{eq : interior region - passes EV - u(y) definition}
    u(y) = {\rm sn}^{-1} \left( \tanh(g), \mu \right) = {\rm sn}^{-1} \left( 1 - \frac{2}{y+1}, \mu \right).
\end{equation}
We can now perform the substitution of $u(y)$ to give
\begin{equation}
    \label{eq : interior region - passes EV - integral (3)}
    \begin{aligned}
        {\rm sgn}(y_R - y_L) \, \frac{2\eta}{\lambda_i}
        &= \frac{1}{|C|} \int_{0}^{u} {\rm cd} \left( t, \mu \right) \, \frac{2 \, {\rm cn}(t, \mu) \, {\rm dn}(t, \mu)}{\left[ 1 - {\rm sn}(t, \mu) \right]^2} dt \\
        &= \frac{2}{|C|} \int_{0}^{u}\frac{1 + 2 \, {\rm sn}(t, \mu) + {\rm sn}^2(t, \mu)}{{\rm cn}^2(t, \mu)} dt \\
        &= \frac{2}{|C|} \int_{0}^{u} \left\{ {\rm nc}^2(t, \mu) + 2 \, \frac{{\rm sn}(t, \mu)}{{\rm cn}^2(t, \mu)} + {\rm sc}^2(t, \mu) \right\} dt \\
        &= \frac{2}{|C|} \left\{ {\rm Nc}(u, \mu) + \frac{2}{1 - \mu^2} \left[ {\rm dc}(u, \mu) - 1 \right] + {\rm Sc}(u, \mu) \right\},
    \end{aligned}
\end{equation}
where ${\rm Nc}$ and ${\rm Sc}$ are defined in Eq.~(\ref{eq : Apx - Jacobi elliptic functions : integral squared naming rule}). Note that the centre integral on the last line can be solved simply by matching terms on the right-hand side of Eq.~(\ref{eq : Apx - Jacobi elliptic functions : derivative of general product}). To neaten the end result slightly, note that by using the identity in Eq.~(\ref{eq : Apx - Jacobi elliptic functions : identity - sn to cn}), we can show that ${\rm Nc}(u, m) - {\rm Sc}(u, m) \equiv u$. Our final result for the field profile is then
\begin{equation}
    \label{eq : interior region - passes EV - field (C > 1)}
    \frac{1}{2 |C|} \left[ u + 2 \, {\rm Sc}(u, \mu) \right] + |C| \left[ {\rm dc}(u, \mu) - 1 \right] = {\rm sgn}(y_R - y_L) \, \frac{(z - z^*)}{2 \lambda_i},
\end{equation}
where $u(y)$ is given in Eq.~(\ref{eq : interior region - passes EV - u(y) definition}). As with the field profile from Section~\ref{sec : Solution for N-layer system - exterior regions}, we cannot invert the above equation into the form $y(z)$. But also as in Section~\ref{sec : Solution for N-layer system - exterior regions} the field has known bounds and is monotonic, so we know $z(y)$ has a one-to-one map to $y(z)$. Next we wish to find the correct values of the constants $C$ and $z^*$ in terms of $y_L$ and $y_R$. Evaluating Eq.~(\ref{eq : interior region - passes EV - field (C > 1)}) at the boundaries and subtracting them to find the difference gives
\begin{equation}
    \label{eq : interior region - passes EV - constraint eq (C > 1)}
    \begin{aligned}
        f_{3}(y_L, y_R, C) := {\rm sgn}(y_R - y_L) \Bigg\{ & \frac{1}{2|C|} \left\{ (u_R - u_L) + 2 \left[ {\rm Sc}(u_R, \mu) - {\rm Sc}(u_L, \mu) \right] \right\} \\
        & + |C| \left[ {\rm dc}(u_R, \mu) - {\rm dc}(u_L, \mu) \right] \Bigg\} = \frac{L_i}{2 \lambda_i},
    \end{aligned}
\end{equation}
where $u_L:=u(y_L)$ and $u_R:=u(y_R)$. We can now numerically solve Eq.~(\ref{eq : interior region - passes EV - constraint eq (C > 1)}) to obtain $\mu$, from which we can compute $z^*$ by evaluating Eq.~(\ref{eq : interior region - passes EV - field (C > 1)}) at either boundary.

Note that if $C^2 < 1$ then $\mu$ becomes imaginary. In this case the field profile in Eq.~(\ref{eq : interior region - passes EV - field (C > 1)}) can be rewritten using Eqs.~(\ref{eq : Apx - Jacobi elliptic functions : argument m is imaginary (sn)}), (\ref{eq : Apx - Jacobi elliptic functions : argument m is imaginary (cn)}), and (\ref{eq : Apx - Jacobi elliptic functions : argument m is imaginary (dn)}) as
\begin{equation}
    \label{eq : interior region - passes EV - field (C < 1)}
    \frac{w}{2} + C^2 \, {\rm Sc}(w, \nu) + |C| \left[ {\rm nc}(w, \nu) - 1 \right] = {\rm sgn}(y_R - y_L) \, \frac{(z - z^*)}{2 \lambda_i},
\end{equation}
where $\nu := \sqrt{1 - C^2}$ and
\begin{equation}
    \label{eq : interior region - passes EV - w(y) definition}
    w(y) = {\rm sd}^{-1} \left( \frac{1}{|C|} \left[ 1 - \frac{2}{y+1} \right], \nu \right).
\end{equation}
Following the same methodology as in the $C^2 \geq 1$ case we find that Eq.~(\ref{eq : interior region - passes EV - constraint eq (C > 1)}) is replaced by
\begin{equation}
    \label{eq : interior region - passes EV - constraint eq (C < 1)}
    \begin{aligned}
        f_{4}(y_L, y_R, C) := {\rm sgn}(y_R - y_L) \Bigg\{ & \frac{(w_R - w_L)}{2} + C^2 \left[ {\rm Sc}(w_R, \nu) - {\rm Sc}(w_L, \nu) \right] \\
        & + |C| \left[ {\rm nc}(w_R, \nu) - {\rm nc}(w_L, \nu) \right] \Bigg\} = \frac{L_i}{2 \lambda_i},
    \end{aligned}
\end{equation}
where $w_L:=w(y_L)$ and $w_R:=w(y_R)$.

\subsection{Junction conditions}
\label{sec : Solution for N-layer system - junction conditions}

In the above section we derived the field profiles for all possible scenarios, with respect to the field values at the boundaries of the region in question. Therefore, all that remains is to find how to solve for these boundary values, given the set of lengths and densities that define the system. To accomplish this we shall use the continuity of the field gradient at each boundary by equating the appropriate pair of Eqs.~(\ref{eq : Exterior regions - first integration}), (\ref{eq : chameleon e.o.m. - integrated once (extrema)}), or (\ref{eq : interior region - passes EV - first integration (non-zero grad at origin)}). In general we integrate Eq.~(\ref{eq : chameleon e.o.m.}) between its boundary and another point within the region with field value $\phi_i$ and field gradient $\phi'_i$. Continuity at the boundary then means
\begin{equation}
    \label{eq : junction conditions - general form - derivation (1)}
    \begin{aligned}
        \frac{1}{2} \left. \left( \frac{d \phi}{dz} \right)^2 \right|_{b,i}
        &= \frac{1}{2} \left( \phi_i' \right)^2 + V^{(i)}_{\rm eff} \left( \phi_{b, i} \right) - V^{(i)}_{\rm eff} \left( \phi_i \right) \\
        &= \frac{1}{2} \left( \phi_{i+1}' \right)^2 + V^{(i+1)}_{\rm eff} \left( \phi_{b, i} \right) - V^{(i+1)}_{\rm eff} \left( \phi_{i+1} \right),
    \end{aligned}
\end{equation}
where $\phi_{b, i}$ is the field value at the boundary joining the two regions. Rearranging this we have
\begin{equation}
\begin{aligned}
    \phi_{b, i} & \frac{(\rho_i - \rho_{i+1})}{M} =
    V^{(i)}_{\rm eff} \left( \phi_{b, i} \right) - V^{(i+1)}_{\rm eff} \left( \phi_{b, i} \right) \\
    &= \frac{1}{2} \left( \phi_{i+1}' \right)^2 - \frac{1}{2} \left( \phi_i' \right)^2 + V^{(i)}_{\rm eff} \left( \phi_i \right) - V^{(i+1)}_{\rm eff} \left( \phi_{i+1} \right) \\
    &= \frac{2 \left( \phi^{\rm EV}_{i+1} \right)^2 C^2_{i+1}}{\lambda^2_{i+1}} - \frac{2 \left( \phi^{\rm EV}_i \right)^2 C^2_i}{\lambda^2_i} + \left[ \frac{\Lambda^5}{\phi_i} + \frac{\rho_i \phi_i}{M} \right] - \left[ \frac{\Lambda^5}{\phi_{i+1}} + \frac{\rho_{i+1} \phi_{i+1}}{M} \right] \\
    &= \frac{4 \rho_{i+1} \phi^{\rm EV}_{i+1} C^2_{i+1}}{M} - \frac{4 \rho_i \phi^{\rm EV}_i C^2_i}{M} + \frac{\rho_i}{M} \left[ \frac{\left( \phi^{\rm EV}_i \right)^2 + \phi_i^2}{\phi_i} \right] - \frac{\rho_{i+1}}{M} \left[ \frac{\left( \phi^{\rm EV}_{i+1} \right)^2 + \phi_{i+1}^2}{\phi_{i+1}} \right] \\
    &= \frac{\rho_i \phi^{\rm EV}_i}{M} \left[ \left( \frac{\left( \phi^{\rm EV}_i \right)^2 + \phi_i^2}{ \phi^{\rm EV}_i \phi_i } \right) - 4 C^2_i \right] - \frac{\rho_{i+1} \phi^{\rm EV}_{i+1}}{M} \left[ \left( \frac{ \left( \phi^{\rm EV}_{i+1} \right)^2 + \phi_{i+1}^2}{ \phi^{\rm EV}_{i+1} \phi_{i+1} } \right) - 4 C^2_{i+1} \right] \\
    &= \frac{\rho_i \phi^{\rm EV}_i}{M} \left[ 2 + \left( \frac{ \left( \phi^{\rm EV}_i - \phi_i\right)^2}{ \phi^{\rm EV}_i \phi_i } \right) - 4 C^2_i \right] - \frac{\rho_{i+1} \phi^{\rm EV}_{i+1}}{M} \left[ 2 + \left( \frac{ \left( \phi^{\rm EV}_{i+1} - \phi_{i+1} \right)^2}{ \phi^{\rm EV}_{i+1} \phi_{i+1} } \right) - 4 C^2_{i+1} \right] \\
    &= 2 \sqrt{\frac{\Lambda^5}{M}} \left( \sqrt{\rho_i} - \sqrt{\rho_{i+1}} \right) \\
    & \qquad + \frac{\rho_i \phi^{\rm EV}_i}{M} \left[ \left( \frac{ \left( \phi^{\rm EV}_i - \phi_i\right)^2}{ \phi^{\rm EV}_i \phi_i } \right) - 4 C^2_i \right] - \frac{\rho_{i+1} \phi^{\rm EV}_{i+1}}{M} \left[ \left( \frac{ \left( \phi^{\rm EV}_{i+1} - \phi_{i+1} \right)^2}{ \phi^{\rm EV}_{i+1} \phi_{i+1} } \right) - 4 C^2_{i+1} \right],
\end{aligned}
\end{equation}
where $C_i := \lambda_i \phi_i' / 2 \phi^{\rm EV}_i$. Therefore, we have that the field value at the boundary of regions $i$ and $i+1$ is
\begin{equation}
    \label{eq : junction conditions - general form}
    \phi_{b, i} = \phi^{HM}_{i, i+1} + \frac{\rho_i \phi^{\rm EV}_i}{(\rho_i - \rho_{i+1})} \delta_i + \frac{\rho_{i+1} \phi^{\rm EV}_{i+1}}{(\rho_{i+1} - \rho_i)} \delta_{i+1},
\end{equation}
where we have introduced the harmonic mean of the field EVs,
\begin{equation}
    \label{eq : junction conditions - harmonic mean field value}
     \phi^{HM}_{i,j} := \frac{\sqrt{\rho_i} \phi^{\rm EV}_i + \sqrt{\rho_j} \phi^{\rm EV}_j}{\sqrt{\rho_i} + \sqrt{\rho_j}} = \frac{2 \sqrt{M \Lambda^5}}{\sqrt{\rho_i} + \sqrt{\rho_j}} = \frac{2}{\big( \phi^{\rm EV}_i \big)^{-1} + \big( \phi^{\rm EV}_j \big)^{-1}},
\end{equation}
and we have used 
\begin{equation}
    \label{eq : junction conditions - general form - general delta (definition)}
    \delta_i := \left[ \left( \frac{ \left( \phi^{\rm EV}_i - \phi_i\right)^2}{ \phi^{\rm EV}_i \phi_i } \right) - 4 C^2_i \right].
\end{equation}
All that is left is to choose appropriate limits of integration for each possible scenario. For exterior regions this will be at $\pm \infty$, while for the interior regions we will choose the origin of our custom coordinates used in each scenario (note if $L_i = L^*_i$ then the origin will lie at either $\pm \infty$). Given this, it is simple to verify that each $\delta_i$ has the form
\begin{equation}
    \label{eq : junction conditions - general form - delta (definition)}
    \delta_i = \left\{ \begin{aligned}
         \ 0 \quad \ \ \ ,& \qquad \text{if $i$ is exterior,} \\
         \ 0 \quad \ \ \ ,& \qquad \text{if $i$ is interior and $L_i = L^*_i$,} \\
        \frac{(1 - k_i)^2}{k_i},& \qquad \text{if $i$ is interior and $L_i > L^*_i$,} \\
        - 4 C_i^2 \quad \ ,& \qquad \text{if $i$ is interior and $L_i < L^*_i$.} 
    \end{aligned} \right.
\end{equation}
Note, in the interior regions the sign of $\delta_i$ identifies the solution scenario.

Our procedure to find the correct set of boundary values starts by defining a universal curve parameter equation for the interior regions by combining Eqs.~(\ref{eq : interior region - maximum - constraint equation (k)}), (\ref{eq : interior region - minimum - constraint equation (k)}), (\ref{eq : interior region - passes EV - constraint eq (C > 1)}), and (\ref{eq : interior region - passes EV - constraint eq (C < 1)}) to give
\begin{equation}
    \label{eq : junction conditions - universal curve parameter equation}
    f^* \left( \phi_L, \phi_R, \gamma \right) := \left\{ \begin{aligned}
        \gamma \, \frac{L_i}{2 \lambda_i} \qquad \qquad, \qquad &\text{if $L_i = L^*_i$}, \\
        f_1 \left( \frac{1}{\gamma} \frac{\phi_L}{\phi^{\rm EV}_i}, \frac{1}{\gamma} \frac{\phi_R}{\phi^{\rm EV}_i}, \gamma \right), \qquad &\text{if $L_i > L^*_i$ and $\phi^{\rm EV}_i > \max \left(\phi_L, \phi_R \right)$}, \\
        f_2 \left( \gamma \frac{\phi_L}{\phi^{\rm EV}_i}, \gamma \frac{\phi_R}{\phi^{\rm EV}_i}, \frac{1}{\gamma} \right), \qquad &\text{if $L_i > L^*_i$ and $\phi^{\rm EV}_i < \min \left(\phi_L, \phi_R \right)$}, \\
        f_3 \left( \frac{\phi_L}{\phi^{\rm EV}_i}, \frac{\phi_R}{\phi^{\rm EV}_i}, \frac{1}{\gamma} - 1 \right), \qquad &\text{if $L_i < L^*_i$ and $\gamma \leq 1/2$}, \\
        f_4 \left( \frac{\phi_L}{\phi^{\rm EV}_i}, \frac{\phi_R}{\phi^{\rm EV}_i}, \frac{1}{\gamma} - 1 \right), \qquad &\text{if $L_i < L^*_i$ and $\gamma > 1/2$},
    \end{aligned} \right.
\end{equation}
where the curve parameter $\gamma \in (0, 1]$ is either $k$ (maximum case), $k^{-1}$ (minimum case), or $1/(1+|C|)$ (EV case). We can then define $\gamma_i \left( \phi_L, \phi_R \right)$ as the solution to $f^* \left( \phi_L, \phi_R, \gamma_i \right) = L_i/2 \lambda_i$ (when $L_i = L^*_i$, then $\gamma_i = 1$ and the field has the same form as the exterior solution in Eq.~(\ref{eq : Exterior regions - field profile})), and plug the result into Eq.~(\ref{eq : junction conditions - general form - delta (definition)}) to obtain $\delta_i \left( \phi_L, \phi_R \right)$. Notice that $\phi_{b, i}$ will have a complicated dependence on $\phi_{b, i\pm1}$ if they belong to interior regions. The problem of solving the field profile in our $N$-layered system is then equivalent to solving the $N-1$ coupled non-linear simultaneous equations for the $\phi_{b, i}$'s. We can accomplish this numerically by solving for each boundary the residual formula
\begin{equation}
    \label{eq : junction conditions - residual formula}
    f_{{\rm residual}, i} \left( \phi_{b,i} \right) = \sigma^{(L)}_i \sqrt{ \left( \frac{\phi_{b, i}}{\phi^{\rm EV}_i} - 1 \right)^2 - \frac{\phi_{b, i}}{\phi^{\rm EV}_i} \, \delta_i} - \sigma^{(R)}_i \sqrt{ \left( \frac{\phi_{b, i}}{\phi^{\rm EV}_{i+1}} - 1 \right)^2 - \frac{\phi_{b, i}}{\phi^{\rm EV}_{i+1}} \, \delta_{i+1}} = 0,
\end{equation}
where $\sigma^{(L)}_i$ and $\sigma^{(R)}_i$ are the signs of the field gradient on the left and right side of the boundary respectively, the values of which are discerned from the scenarios in the two regions joined at the boundary (see Eqs.~(\ref{eq : Exterior regions - field gradient}), (\ref{eq : interior region - field gradient (maximum)}), (\ref{eq : interior region - field gradient (minimum)}), and (\ref{eq : interior region - passes EV - field gradient})). This residual was derived from Eq.~(\ref{eq : junction conditions - general form - derivation (1)}), and contains the same information as Eq.~(\ref{eq : junction conditions - general form}), while also enforcing that the field gradient has the same sign on either side of the boundary. The $\delta_i$'s are computed from Eqs.~(\ref{eq : junction conditions - general form - delta (definition)}) and (\ref{eq : junction conditions - universal curve parameter equation}).

\section{Example Solutions}
\label{sec : Example Solutions}

\subsection{Single Plate}
\label{sec : Example Solutions - Single Plate}

\begin{figure}[ht]
    \centering
    \includegraphics[width=1.0\linewidth]{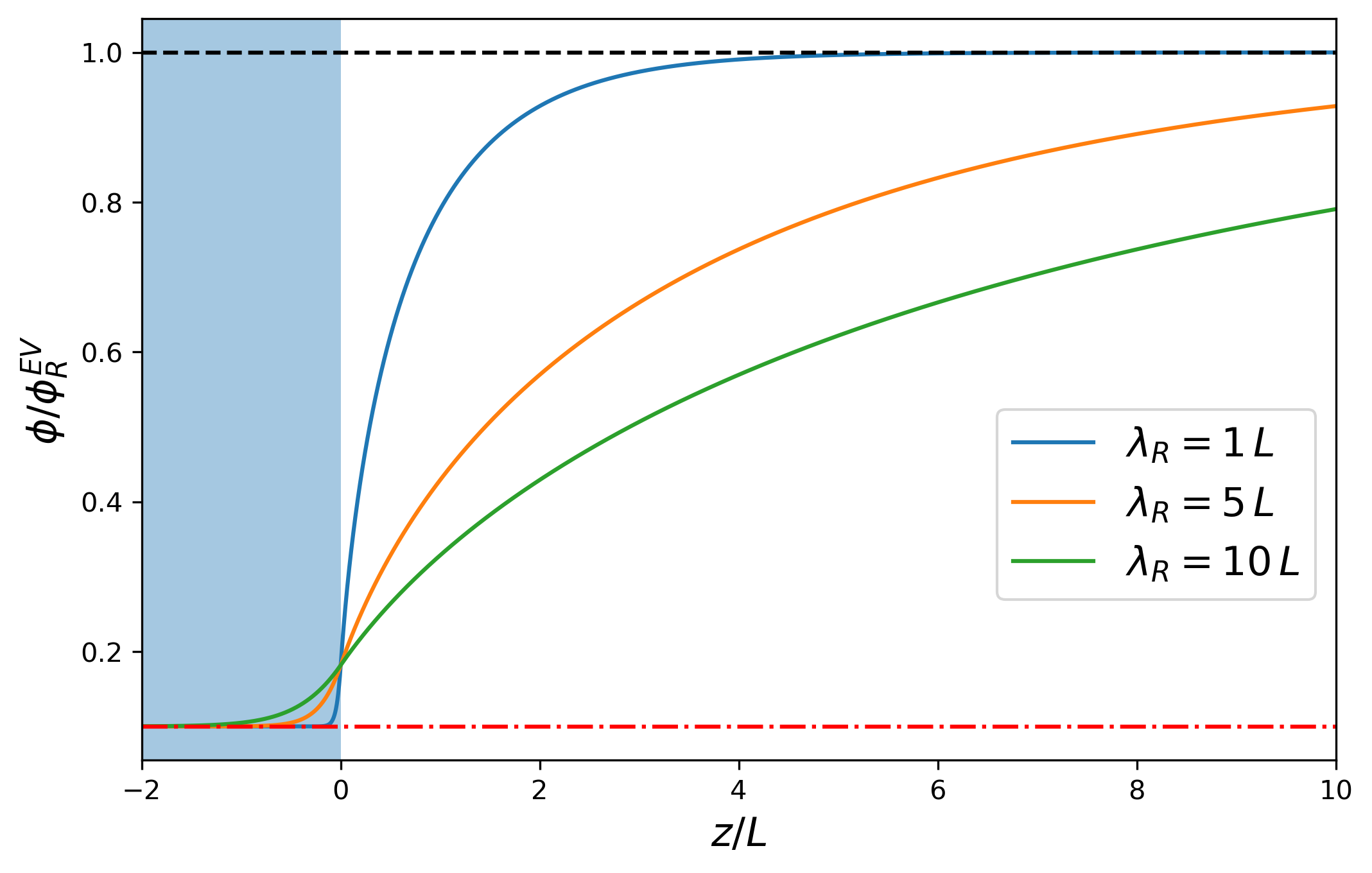}
    \caption{Field profiles for the $2$-layered system where the left region (shaded in blue) has a density $\rho_L = 100\, \rho_R$, and we are using an arbitrary length scale $L$. Each curve corresponds to different values of Compton wavelengths in the right region (see Eq.~(\ref{eq : chameleon Compton wavelength})). The horizontal red dash-dotted and black dashed lines indicate the values of the field EV (obtained using Eq.~(\ref{eq : expectation Value (EV)})) inside the left and right regions respectively.}
    \label{fig : Examples - two-layers - field}
\end{figure}

In this case we have two layers with densities $\rho_L$ and $\rho_R$ (respectively $\rho_1$ and $\rho_2$), with a shared boundary at $z=0$. Using Eqs.~(\ref{eq : junction conditions - general form}) and (\ref{eq : junction conditions - general form - delta (definition)}) we see that the field value at the boundary of the two regions is simply the harmonic mean field value defined in Eq.~(\ref{eq : junction conditions - harmonic mean field value}). Recall that the field profile in the exterior region is given by Eq.~(\ref{eq : Exterior regions - field profile}), which has $\phi_b$ as its only unknown. Therefore, we have fully determined the field profile everywhere (see Fig.~\ref{fig : Examples - two-layers - field} for examples with different density ratios).

This scenario could represent a single plate of dense material within a partial vacuum. Such a setup was used in~\cite{brax2011strongly}, in which the effect of the chameleon on the energy levels of a neutron bouncing on a mirror was calculated. We note that the solution used in~\cite{brax2011strongly} is not exact, as it assumes $\phi/M \ll 1$ and $\phi \ll \phi^{\rm EV}_R$. In Appendix~\ref{sec : Apx - Matching to known approximate analytic solutions} we show that by making the same assumptions, and applying them to Eq.~(\ref{eq : Exterior regions - field profile}), we reproduce their solution for the field profile.

Since this solution requires two infinitely extended regions, it cannot be realised in reality. That said, in cases where the field's Compton wavelength is sufficiently small in both regions when compared to the distance to any other objects, i.e. changes in density, this solution then becomes a good approximation. This limits the usefulness of this setup. Consequently, most of the literature regarding planar chameleon solutions uses a two plate setup (3-layered), which we will discuss next.

\subsection{Pair of Plates}
\label{sec : Example Solutions - Pair of Plates}

In this section we will consider the $3$-layer case of a partial vacuum of density $\rho_V$ between a pair of parallel plates, separated by distance $L$, with densities $\rho_L$ and $\rho_R$ for the left and right plates respectively ($\rho_V \ll \rho_L, \rho_R$). This type of setup has been used extensively in the literature, including: atom interferometry~\cite{burrage2016proposed, sabulsky2019experiment}, neutron interferometry~\cite{lemmel2015neutron}, Casimir force experiments~\cite{brax2010tuning, almasi2015force}, and neutron bouncing experiments~\cite{brax2013probing}, such as \textit{q}BOUNCE~\cite{ivanov2013influence}. However, in each of these works the field profile used was an approximation of the true solution, obtained by taking the field's Compton wavelength to be sufficiently small~\cite{burrage2016proposed, brax2010tuning} or the vacuum to be perfect~\cite{ivanov2013influence}.

\begin{figure}
    \centering
    \includegraphics[width=1.0\linewidth]{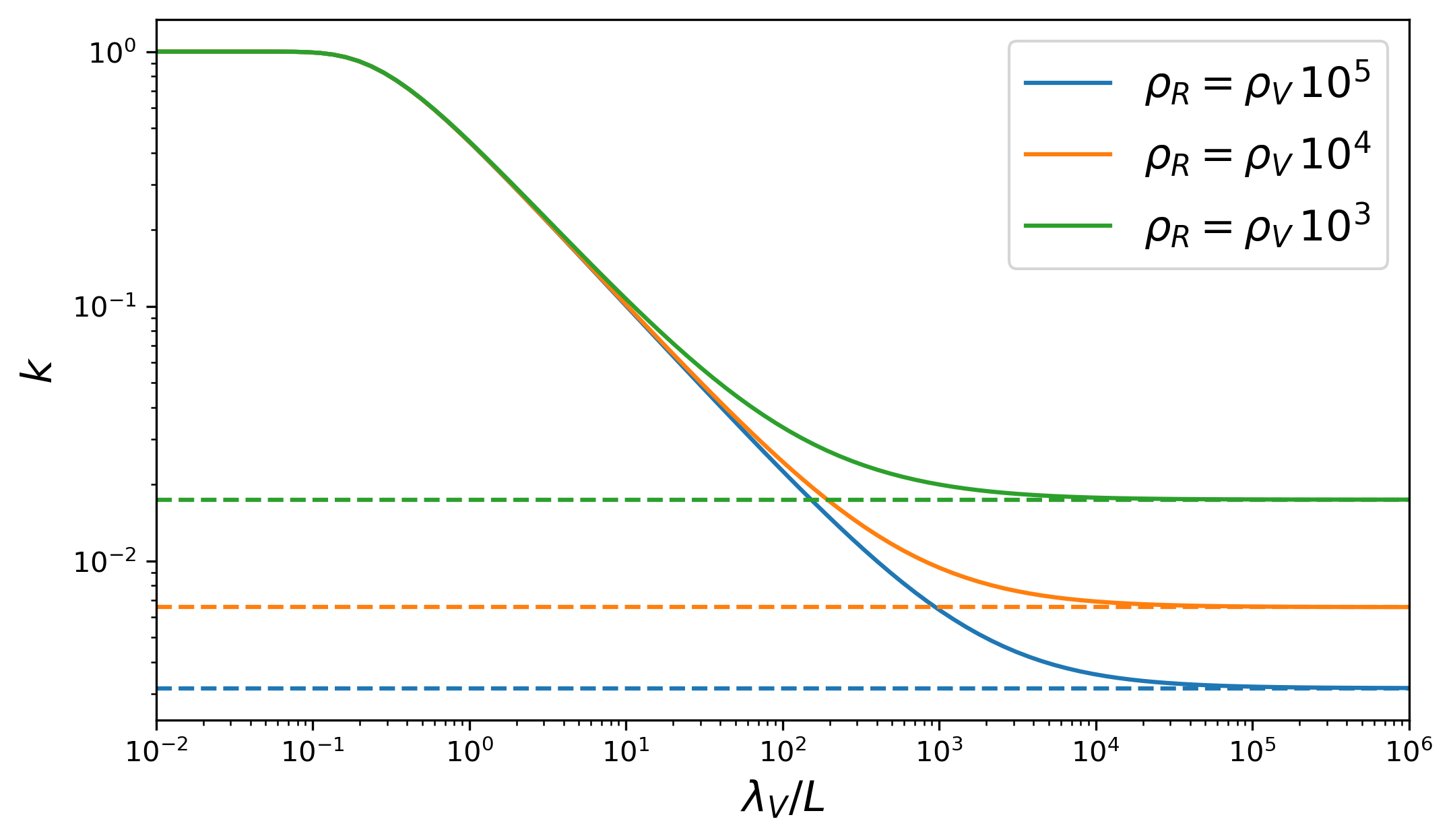}
    \caption{Distribution of solutions to Eq.~(\ref{eq : interior region - maximum - constraint equation (k)}) on the $\lambda_V$-$k$ plane. In each curve we vary the value of $\rho_R$ while keeping $\rho_L=10^5\,\rho_V$. The dashed lines indicate the values of $k_{\rm min}$ predicted by Eq.~(\ref{eq : Example Solutions - three layers - k limit for lambda_V --> infinity}).}
    \label{fig : Examples - three-layers - k values}
\end{figure}

\begin{figure}[ht]
    \centering
    \includegraphics[width=1.0\linewidth]{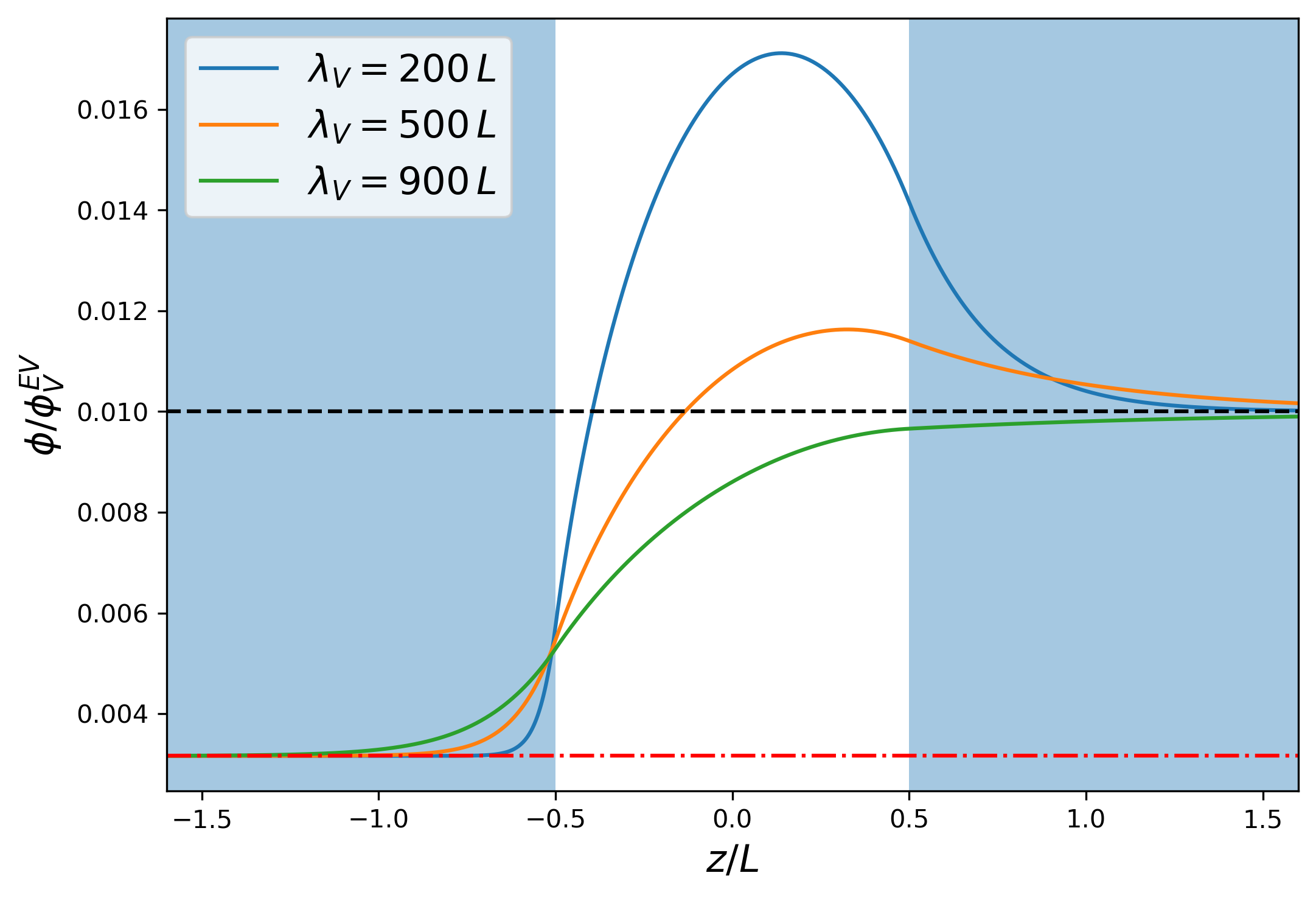}
    \caption{Plot of analytic solutions for multiple values of the vacuum Compton wavelength. The shaded regions represent the plates, with densities $10^5$ (left) and $10^4$ (right) times larger than the vacuum. The red dash-dotted and black dashed lines indicate the EVs inside the left and right plate respectively.}
    \label{fig : Examples - three-layers - field}
\end{figure}

We shall define our origin to be in the centre of the vacuum, while the boundaries of the plates are located at $z=\pm L/2$. We can also orientate our coordinate so $\rho_L \geq \rho_R$. Since we only have one interior region we will not have any interior-interior boundaries. If we consider the extremum scenario, Eq.~(\ref{eq : junction conditions - general form}) becomes
\begin{equation}
    \label{eq : Example Solutions - three layers - field at walls (extrema)}
    x_j = 1 - \frac{(k - \omega_j)^2}{k^2 (1 - \omega_j^2)},
\end{equation}
where $j \in \{L, R\}$, $\omega_j := \phi^{\rm EV}_j/\phi^{\rm EV}_V$, and $x_j := \phi_j/\phi_0$. Plugging this into Eq.~(\ref{eq : interior region - maximum - constraint equation (k)}) then gives us our constraint equation for $k$. The relation between $k$ and the Compton wavelength in the vacuum is shown in Fig.~\ref{fig : Examples - three-layers - k values}, while the field profiles are shown in Fig.~\ref{fig : Examples - three-layers - field}. The plateaus on the right of Fig.~\ref{fig : Examples - three-layers - k values} can be derived by taking the limit $\lambda_V \rightarrow \infty$ and applying this to Eqs.~(\ref{eq : interior region - maximum - constraint equation (k)}) and (\ref{eq : Example Solutions - three layers - field at walls (extrema)}). In the case when $\rho_L = \rho_R$, symmetry dictates that the maximum must lie inside the vacuum region, and so Eq.~(\ref{eq : interior region - maximum - constraint equation (k)}) becomes $x_L = x_R = 1$, and Eq.~(\ref{eq : Example Solutions - three layers - field at walls (extrema)}) then becomes $k_{\rm min} = \omega_L = \omega_R$. Meanwhile, in the $\rho_L \neq \rho_R$ scenario the maximum must lie outside the vacuum region since Eq.~(\ref{eq : interior region - maximum - monotonic constraint}) becomes $x_c = 1$, and so the field is monotonic in the vacuum. In this case Eq.~(\ref{eq : interior region - maximum - constraint equation (k)}) becomes $x_L = x_R$, which if applied to Eq.~(\ref{eq : Example Solutions - three layers - field at walls (extrema)}) gives us the expression for the value of $k$ as $\lambda_V \rightarrow \infty$ as
\begin{equation}
    \label{eq : Example Solutions - three layers - k limit for lambda_V --> infinity}
    k_{\rm min} = \frac{\omega_R \, \sqrt{1 - \omega_L^2} + \omega_L \, \sqrt{1 - \omega_R^2}}{\sqrt{1 - \omega_R^2} + \sqrt{1 - \omega_L^2}}.
\end{equation}
Notice if $\omega_L = \omega_R$ then this expression reduces to that of the $\rho_L = \rho_R$ scenario. Therefore, Eq.~(\ref{eq : Example Solutions - three layers - k limit for lambda_V --> infinity}) holds for all combinations of $\rho_L$ and $\rho_R$. We can also see that we will remain in the extremum scenario for arbitrarily large $\lambda_V$, since $k \rightarrow k_{\rm min} < 1$.

To conclude this sub-section, we direct the reader to Appendix~\ref{sec : Apx - Matching to known approximate analytic solutions} where the solution presented above is shown to reproduce the approximate solutions for the parallel plate scenario found in~\cite{ivanov2013influence, burrage2016proposed}.

\subsection{Casimir force experiments and model constraints}
\label{sec : Example Solutions - Casimir experiments and model constraints}

\begin{figure}[ht]
    \centering
    \includegraphics[width=1.0\linewidth]{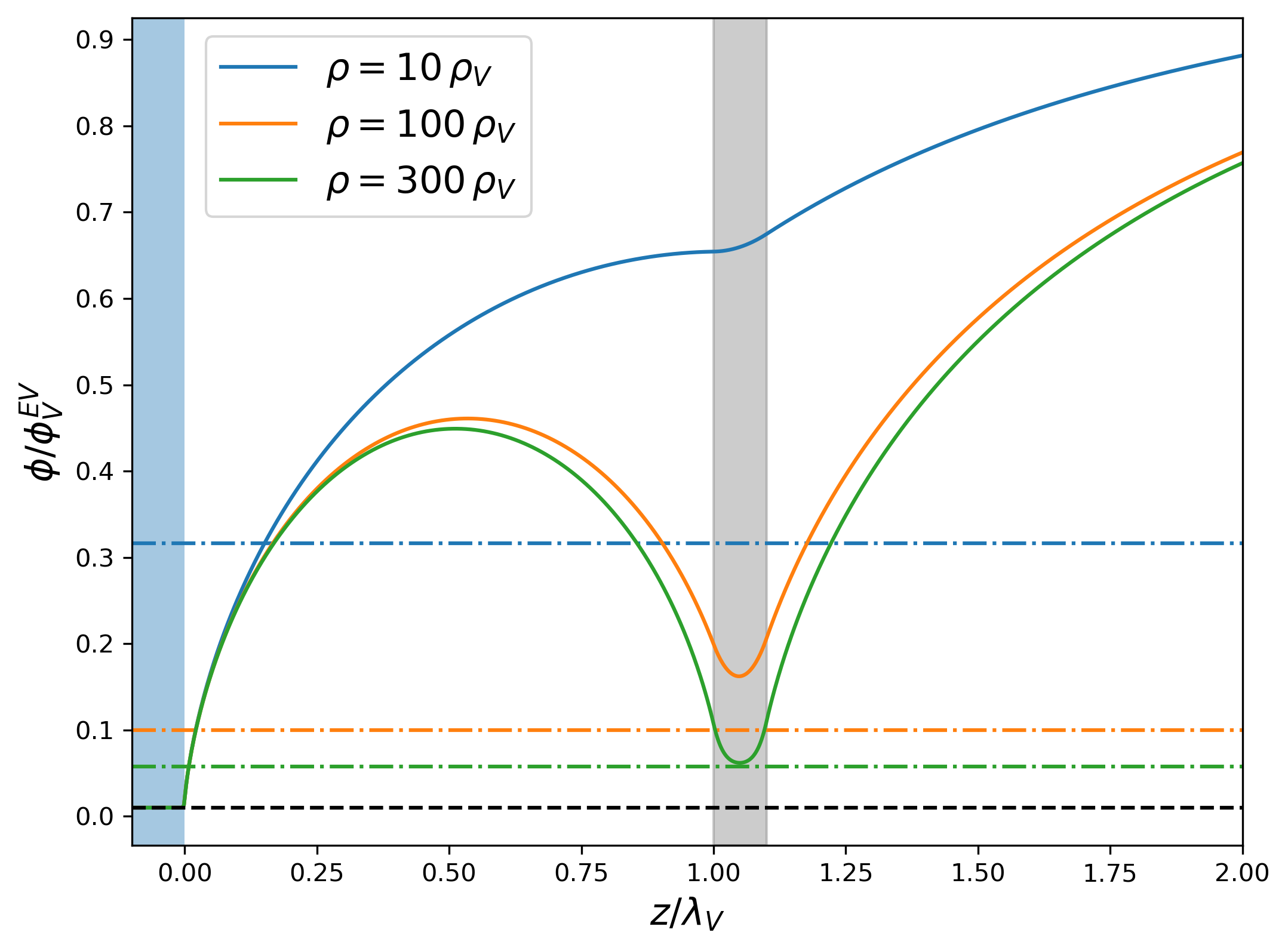}
    \caption{Field profiles for a Casimir-type experiment consisting of a wall, and a plate surrounded by a partial vacuum. The wall (blue shaded region) has a density $10^4$ times larger than the vacuum (in white) and is taken to be infinitely extended in the negative $z$-direction. The plate (shaded in grey) is located a distance $\lambda_V$ (the vacuum Compton wavelength) from the wall and has a thickness of $0.1\, \lambda_V$. The field is plotted for a range of plate densities, with the corresponding EV of the field in the plate shown by a dash-dotted line of the corresponding colour. The black dashed line shows the EV inside the wall. The vacuum to the right of the plate is extended infinitely in the positive $z$-direction.}
    \label{fig : Examples - Four-layers - vary plate}
\end{figure}

\begin{figure}[ht]
    \centering
    \includegraphics[width=1.0\linewidth]{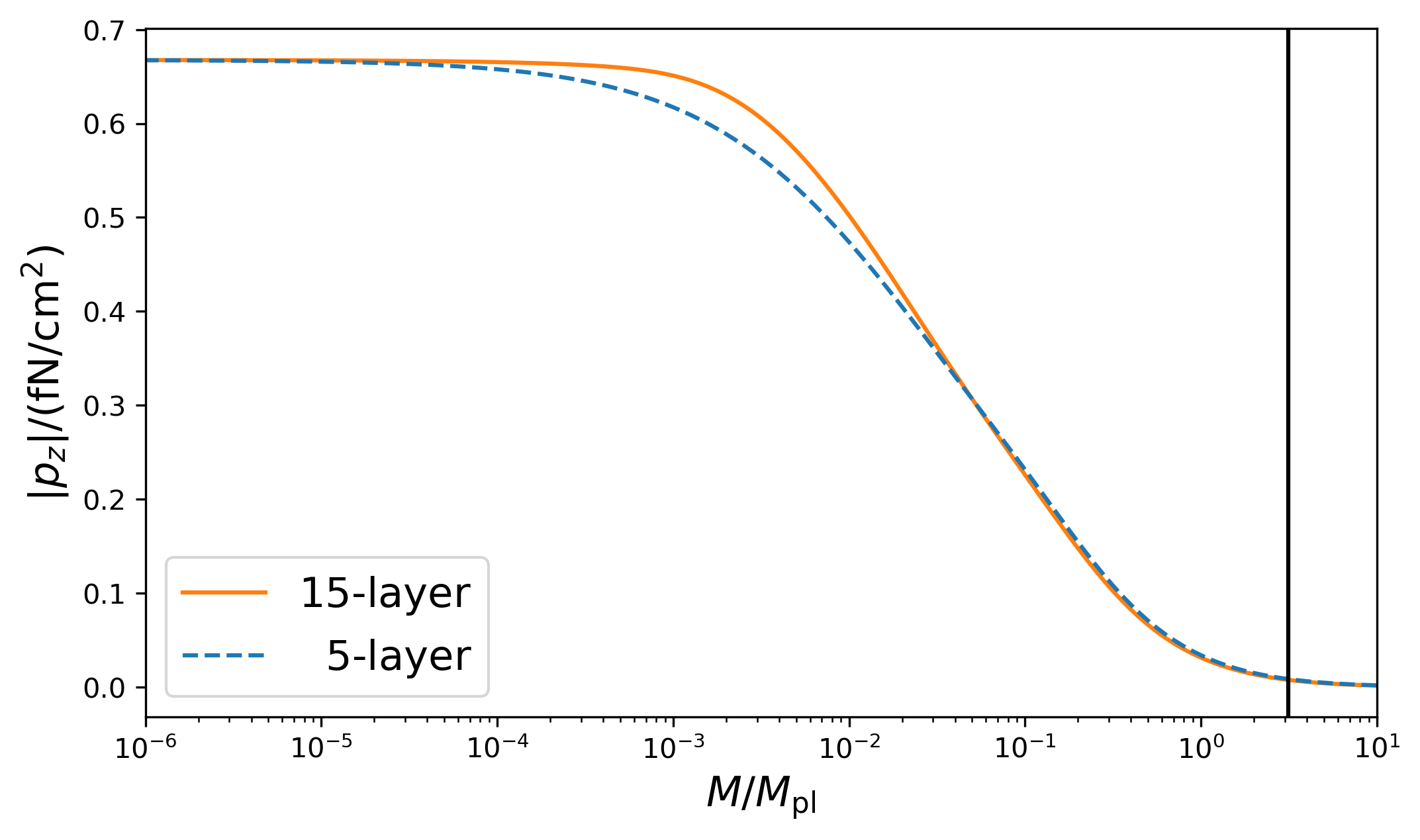}
    \caption{Plot of the chameleon pressure, as a function of $M$ and with $\Lambda = 2.4 \, {\rm meV}$, induced on the sensor plate in \textsc{Cannex}. The solid orange line is the curve for the exact setup consisting of fifteen layers, while the dashed blue line is a $5$-layer approximation of \textsc{Cannex} where the walls and sensor plate have uniform densities (using the respective weighted averages for the densities). The size of the vacuum region in both cases is $20.4 \, \rm{\mu m}$. The vertical black line shows when the field's Compton wavelength inside the aluminium shielding is $0.3 \, {\rm mm}$, which is one-tenth the thickness of the shield. Larger Compton wavelengths may lead the true field profile to deviate from our analytic solution, since the finite thickness of the outermost layer can then no longer be neglected.}
    \label{fig : Examples - Cannex - pressure vs M}
\end{figure}

Finally, we consider Casimir-type experiments consisting of a pair of plates, which are separated by a low-density region. The plates suppress the field values of the hypothetical chameleon, thereby inducing an effective pressure on the plates. The simplest model of a Casimir force experiment is a $4$-layer scenario, as depicted in Fig.~\ref{fig : Examples - Four-layers - vary plate}, comprising of two plates: a fixed plate and a movable sensor plate. The induced pressure on the sensor plate is given by~\cite{MarioHabil}
\begin{equation}
    \label{eq : Example Solutions - Casimir - pressure (general)}
    p = - \frac{1}{M} \int_{z_L}^{z_R} \rho(z) \, \partial_z\phi(z) \, dz,
\end{equation}
which for a plate of constant density, $\rho_M$, reduces to
\begin{equation}
    \label{eq : Example Solutions - Casimir - pressure (constant density)}
    p = \frac{\rho_M}{M} \left(\phi_L - \phi_R \right),
\end{equation}
where $\phi_L$ and $\phi_R$ are the field values on the left and right boundaries of the sensor plate respectively. By measuring the induced pressure on the sensor plate and accounting for the background pressure due to the Casimir effect, we can detect the chameleon field or constrain its parameter values. One such experiment capable of this is \textsc{Cannex}, for which our analytic solution can be employed to obtain the corresponding constraint curves. However, the plates used in \textsc{Cannex} are composed of several layers of different materials and densities, and hence the corresponding setup is more complicated than the $4$-layer system depicted in Fig.~\ref{fig : Examples - Four-layers - vary plate} (details of the layer structure and sensitivity were obtained both from~\cite{Haghmoradi:2024ytw} and via private communication). In this more realistic case Eq.~(\ref{eq : Example Solutions - Casimir - pressure (general)}) becomes
\begin{equation}
    \label{eq : Example Solutions - Casimir - pressure (Cannex)}
    p = \frac{1}{M} \left\{ \rho_1 \left( \phi_0 - \phi_1 \right) + \rho_2 \left( \phi_1 - \phi_2 \right) + ... + \rho_N \left( \phi_{N-1} - \phi_N \right) \right\},
\end{equation}
where from left-to-right the sensor plate's sub-layers have densities $\{\rho_1, \rho_2, ..., \rho_N\}$, with boundary values $\{\phi_0, \phi_1, ..., \phi_N\}$. Therefore, after finding all the boundary values using the method described in Section~\ref{sec : Solution for N-layer system - junction conditions}, we can compute from Eq.~(\ref{eq : Example Solutions - Casimir - pressure (Cannex)}) the chameleon pressure experienced by the sensor plate. The strength of this pressure is shown in Fig.~\ref{fig : Examples - Cannex - pressure vs M}, along with the pressure obtained if we had approximated \textsc{Cannex} by taking the wall and plate to have uniform densities. As is to be expected at large $M$, which using Eq.~(\ref{eq : chameleon Compton wavelength}) corresponds to larger Compton wavelength, the field cannot resolve the small sub-layers, resulting in an effective smearing out of the density profile. Therefore, in this limit it is valid to approximate the true setup with a $5$-layer (this includes a second wall to contain the vacuum) counterpart with the wall and plate densities being given by the weighted average density of their respective sub-layers. Meanwhile at small $M$, and smaller Compton wavelength, the plate and wall are screened so that only the outermost layer contributes to the field in the vacuum regions, and again the $5$-layer approximation is valid. We also see how the deviation becomes larger in the intermediate region, with its maximum being $\sim 7 \%$ larger than the approximated $5$-layer case.

\begin{figure}
    \centering
    \includegraphics[width=1.0\linewidth]{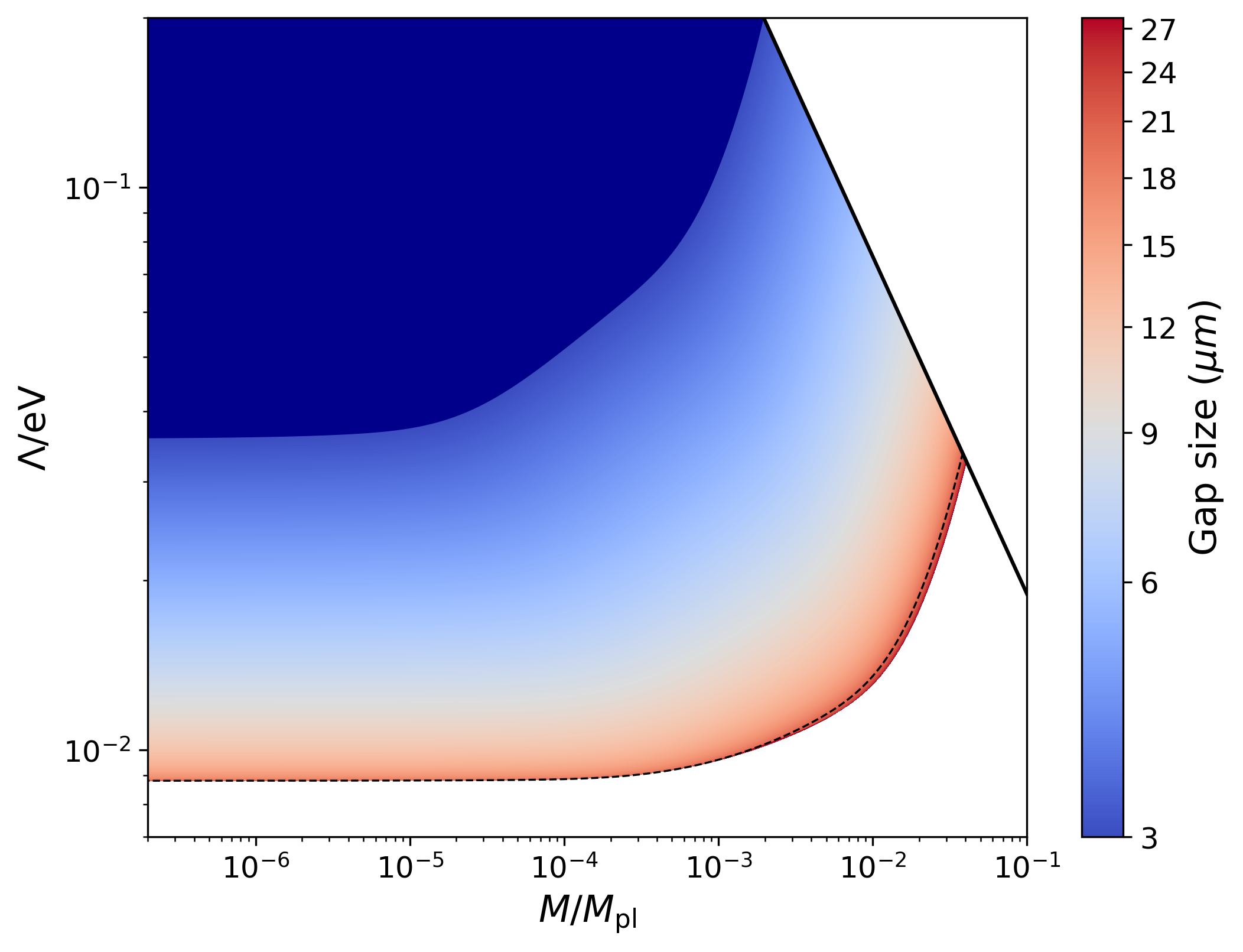}
    \caption{Forecasted constraints for \textsc{Cannex} ($95 \%$ C.L.). The heat map shows the minimum vacuum gap size required to constrain that region of the parameter space (the solid blue region is constrained for all gap sizes). Note that the pressure sensitivity varies with the vacuum gap size. The solid black line indicates when the field's Compton wavelength inside the outer wall becomes one-tenth of the wall's width. Beyond this we would expect our solution to no longer be a good approximation, since it assumes the outer layer to be infinite in length. The dashed black line indicates the bound given by a vacuum gap size of $20.4 \, \rm{\mu m}$.}
    \label{fig : Example - Cannex - Constraint plot (gap size)}
\end{figure}

\begin{figure}
    \centering
    \includegraphics[width=1.0\linewidth]{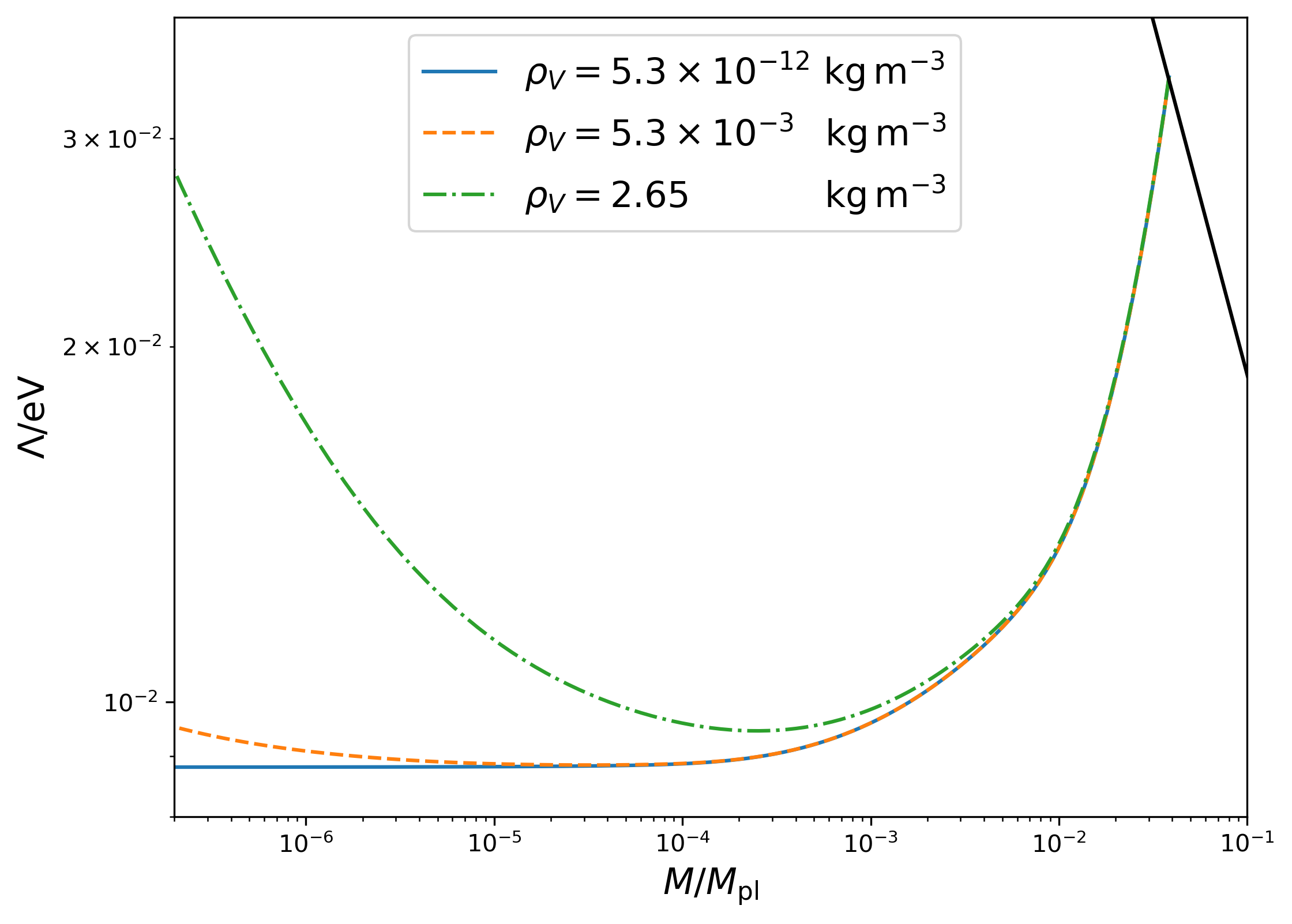}
    \caption{Lower bounds of the region constrained by \textsc{Cannex} for different gas densities inside the vacuum gap ($95 \%$ C.L.). The stated densities correspond to Xe gas at $25\,^\circ\text{C}$ and pressures of $10^{-9} \, {\rm mbar}$ (blue), $1 \, {\rm mbar}$ (orange), and $500 \, {\rm mbar}$ (green). We assumed a vacuum gap size of $20.4 \, \rm{\mu m}$, and that the pressure sensitivity does not change. The solid black line represents when the field's Compton wavelength inside the outer wall becomes one-tenth of the wall's width, beyond which the solution may become untrustworthy.}
    \label{fig : Example - Cannex - Constraint plot (vacuum density)}
\end{figure}

\begin{figure}
    \centering
    \includegraphics[width=1.0\linewidth]{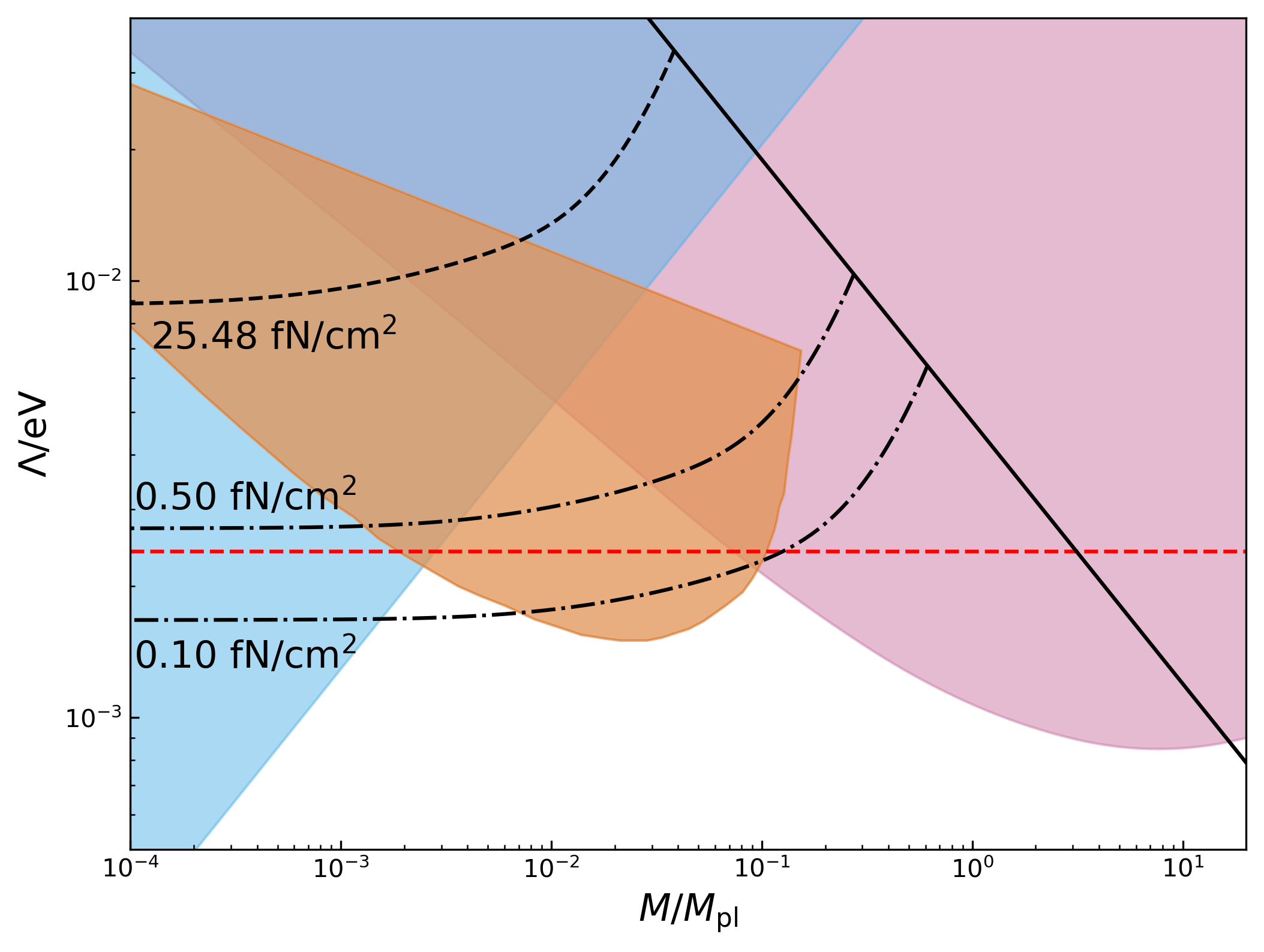}
    \caption{Constraint plot for the chameleon ($n=1$) model, with theoretical forecasts from \textsc{Cannex} with a vacuum gap of $20.4 \, \rm{\mu m}$ ($95 \%$ C.L.). The lower bound of the constrained region is indicated by the dashed line (with pressure sensitivity of $25.48 \, \rm{fN/cm^2}$), while the dash-dotted lines indicate the same curve at higher sensitivities. The shaded regions represent existing constraints from torsion pendulum~\cite{upadhye2012dark} (purple), atom interferometry~\cite{mueller2016atom} (blue), and levitated force sensor~\cite{yin2022experiments} (orange). We choose to apply an artificial cutoff (indicated by the solid black line) where the field's Compton wavelength inside the aluminium shielding is equal to $0.3 \, {\rm mm}$, which is one-tenth the thickness of the shield. Beyond this cutoff, boundary effects due to our outer layer having a finite length may start to become non-negligible. The dashed red line indicates the dark energy scale of $\Lambda = 2.4 \, {\rm meV}$.}
    \label{fig : Examples - Cannex - Constraint plot}
\end{figure}

Comparing the calculated chameleon pressure with the pressure sensitivity of \textsc{Cannex} (for $95 \%$ C.L. we require $|p(d)| > 2\sigma(d)$, where $\sigma$ is the sensitivity and $d$ is the vacuum gap size) allows us to forecast the resulting model constraints. Within \textsc{Cannex} both the vacuum density and gap size can be varied. Using our solution, we show how doing so would affect the forecasted constraints. These are plotted in Fig.~\ref{fig : Example - Cannex - Constraint plot (gap size)} and Fig.~\ref{fig : Example - Cannex - Constraint plot (vacuum density)} for the gap size and density respectively. We find that, to reach lower values of $\Lambda$, the ideal vacuum gap size is $20.4 \,\rm{\mu m}$, while the density of the gas in the vacuum gap should be as low as possible. We compare our forecast for \textsc{Cannex}, with a vacuum gap size of $20.4 \, \rm{\mu m}$, to the existing constraints in Fig.~\ref{fig : Examples - Cannex - Constraint plot}. At this gap size the pressure sensitivity is $25.48 \, \rm{fN/cm^2}$, which is too large to contribute new constraints for the $n=1$ chameleon model (comments regarding $n>1$ models are beyond the scope of this work).

\section{Conclusion}

In this work we have derived the equations and general methodology required to find exact solutions for the $n=1$ chameleon field for systems of $N$ discrete planar layers. The only assumptions made in this work are that the discrete layers are perfectly uniform, and that the outermost layers are infinitely extended. While the former is a reasonable idealisation, the latter is obviously unphysical. Therefore, our solutions are only valid when the thickness of the outer layers is significantly larger than the corresponding Compton wavelengths, so that from the field's perspective the outer layers are, to a good approximation, infinitely thick. This will correspond to when the field approximately reaches its respective EVs in each outer layer. That said, our solutions are applicable to a range of experiments including E$\rm\ddot{o}$t-Wash~\cite{upadhye2012dark}, \textsc{Cannex}~\cite{sedmik2021next, Haghmoradi:2024ytw} and \textit{q}BOUNCE~\cite{ivanov2013influence, jenke2014gravity}. Using \textsc{Cannex} as an example, we have illustrated the ability to forecast model constraints using our solutions. We have also shown how the results would differ between the true experimental setup and an approximation of it. Although in this case the difference is small, it is important to consider this when designing future experiments. Finally, we have shown in Appendix~\ref{sec : Apx - Matching to known approximate analytic solutions} that our exact solutions agree with the approximate solutions that already exist in the literature. The solution in~\cite{ivanov2016exact} was found to have multiple mistakes in its methodology and so is examined separately in Appendix~\ref{sec : Apx - Correcting previous derivation}.

\FloatBarrier
\section*{Acknowledgments} 
This research was funded in whole or in part by the Austrian Science Fund (FWF) 
[10.55776/PAT7599423]. For open access purposes, the authors have applied a CC BY public copyright license to any author accepted manuscript version arising from this submission. The authors acknowledge TU Wien Bibliothek for financial support through its Open Access Funding Programme. We would also like to thank Hartmut Abele and René Sedmik for useful discussions. In addition, we honour the memory of Andrey Nikolaevich Ivanov, who sadly passed away in 2021. Andrey has been a guest professor at the Atominstitut and a close collaborator of one of the authors (M.P.) for many years.

\section*{Data and code availability}
All code and data used in this work can be found at \url{https://github.com/C-Briddon/chameleon_n1_planar_solutions}. This includes a Python package that uses the solutions derived above to construct the field profile for an $N$-layered system.

\bibliographystyle{unsrt}
\bibliography{References.bib}{}

@article{ivanov2016exact,
  author  = {Ivanov, A. N. and Cronenberg, G. and H{\"o}llwieser, R. and Jenke, T. and Pitschmann, M. and Wellenzohn, M. and Abele, H.},
  title   = {Exact solution for chameleon field, self-coupled through the {Ratra-Peebles} potential with $n = 1$ and confined between two parallel plates},
  journal = {Physical Review D},
  volume  = {94},
  number  = {8},
  pages   = {085005},
  year    = {2016},
  doi     = {10.1103/PhysRevD.94.085005}
}

@article{briddon2021selcie,
  author    = {Briddon, C. and Burrage, C. and Moss, A. and Tamosiunas, A.},
  title     = {{SELCIE}: a tool for investigating the chameleon field of arbitrary sources},
  journal   = {Journal of Cosmology and Astroparticle Physics},
  volume    = {2021},
  number    = {12},
  pages     = {043},
  year      = {2021},
  doi       = {10.1088/1475-7516/2021/12/043},
  publisher ={IOP Publishing}
}

@phdthesis{MarioHabil,
  Type   = {Habilitation thesis},
  Title  = "{The High Precision Frontier: Search for New Physics with “Tabletop Experiments” \& Beyond}",
  Author = {Pitschmann, M.},
  School = {TU Wien},
  Year   = {2023},
}

@book{abramowitz1964handbook,
  editor    = {Abramowitz, M. and Stegun, I. A.},
  title     = {Handbook of Mathematical Functions with Formulas, Graphs, and Mathematical Tables},
  series    = {National Bureau of Standards Applied Mathematics Series},
  volume    = {55},
  publisher = {U.S. Government Printing Office},
  address   = {Washington, D.C.},
  year      = {1964}
}

@article{sabulsky2019experiment,
  author  = {Sabulsky, D. O. and Dutta, I. and Hinds, E. A. and Elder, B. and Burrage, C. and Copeland, E. J.},
  title   = {Experiment to detect dark energy forces using atom interferometry},
  journal = {Physical Review Letters},
  volume  = {123},
  number  = {6},
  pages   = {061102},
  year    = {2019},
  doi     = {10.1103/PhysRevLett.123.061102}
}

@article{fischer2024screened,
  author  = {Fischer, H. and K{\"a}ding, C. and Pitschmann, M.},
  title   = {Screened scalar fields in the laboratory and the solar system},
  journal = {Universe},
  volume  = {10},
  number  = {7},
  pages   = {297},
  year    = {2024},
  doi     = {10.3390/universe10070297}
}

@article{fischer2024numerical,
  title={Numerical methods for scalar field dark energy in tabletop experiments and Lunar Laser Ranging},
  author={Fischer, H. and Sedmik, R. I. P.},
  journal={Journal of Cosmology and Astroparticle Physics},
  volume={2024},
  number={10},
  pages={026},
  year={2024},
  publisher={IOP Publishing}
}

@article{ivanov2013influence,
  author  = {Ivanov, A. N. and H{\"o}llwieser, R. and Jenke, T. and Wellenzohn, M. and Abele, H.},
  title   = {Influence of the chameleon field potential on transition frequencies of gravitationally bound quantum states of ultracold neutrons},
  journal = {Physical Review D},
  volume  = {87},
  number  = {10},
  pages   = {105013},
  year    = {2013},
  doi     = {10.1103/PhysRevD.87.105013}
}

@article{brax2011strongly,
  author  = {Brax, P. and Pignol, G.},
  title   = {Strongly coupled chameleons and the neutronic quantum bouncer},
  journal = {Physical Review Letters},
  volume  = {107},
  number  = {11},
  pages   = {111301},
  year    = {2011},
  doi     = {10.1103/PhysRevLett.107.111301}
}

@article{brax2013probing,
  author        = {Brax, P. and Pignol, G. and Roulier, D.},
  title         = {Probing strongly coupled chameleons with slow neutrons},
  journal       = {Physical Review D},
  volume        = {88},
  number        = {8},
  pages         = {083004},
  year          = {2013},
  doi           = {10.1103/PhysRevD.88.083004},
  eprint        = {1306.6536},
  archivePrefix = {arXiv}
}

@article{burrage2016proposed,
  author  = {Burrage, C. and Copeland, E. J. and Stevenson, J. A.},
  title   = {A proposed experimental search for chameleons using asymmetric parallel plates},
  journal = {Journal of Cosmology and Astroparticle Physics},
  volume  = {2016},
  number  = {08},
  pages   = {070},
  year    = {2016},
  doi     = {10.1088/1475-7516/2016/08/070}
}

@article{lemmel2015neutron,
  author  = {Lemmel, H. and Brax, P. and Ivanov, A. N. and Jenke, T. and Pignol, G. and Pitschmann, M. and Potocar, T. and Wellenzohn, M. and Zawisky, M. and Abele, H.},
  title   = {Neutron interferometry constrains dark energy chameleon fields},
  journal = {Physics Letters B},
  volume  = {743},
  pages   = {310--314},
  year    = {2015}
}

@article{brax2010tuning,
  author  = {Brax, Ph. and van de Bruck, C. and Davis, A.-C. and Shaw, D. J. and Iannuzzi, D.},
  title   = {Tuning the mass of chameleon fields in {Casimir} force experiments},
  journal = {Physical Review Letters},
  volume  = {104},
  number  = {24},
  pages   = {241101},
  year    = {2010},
  doi     = {10.1103/PhysRevLett.104.241101}
}

@article{sedmik2021next,
  author  = {Sedmik, R. I. P. and Pitschmann, M.},
  title   = {Next generation design and prospects for {CANNEX}},
  journal = {Universe},
  volume  = {7},
  number  = {7},
  pages   = {234},
  year    = {2021},
  doi     = {10.3390/universe7070234}
}

@article{Haghmoradi:2024ytw,
  author  = {Haghmoradi, H. and Fischer, H. and Bertolini, A. and Gali{\'c}, I. and Intravaia, F. and Pitschmann, M. and Schimpl, R. A. and Sedmik, R. I. P.},
  title   = {Force metrology with plane parallel plates: {F}inal design review and outlook},
  journal = {Physics},
  volume  = {6},
  number  = {2},
  pages   = {690--741},
  year    = {2024},
  doi     = {10.3390/physics6020045}
}

@article{jenke2014gravity,
  author  = {Jenke, T. and Cronenberg, G. and Burgd{\"o}rfer, J. and Chizhova, L. A. and Geltenbort, P. and Ivanov, A. N. and Lauer, T. and Lins, T. and Rotter, S. and Saul, H. and others},
  title   = {Gravity resonance spectroscopy constrains dark energy and dark matter scenarios},
  journal = {Physical Review Letters},
  volume  = {112},
  number  = {15},
  pages   = {151105},
  year    = {2014},
  doi     = {10.1103/PhysRevLett.112.151105}
}

@article{mueller2016atom,
  author        = {Elder, B. and Khoury, J. and Haslinger, P. and Jaffe, M. and M{\"u}ller, H. and Hamilton, P.},
  title         = {Chameleon dark energy and atom interferometry},
  journal       = {Physical Review D},
  volume        = {94},
  number        = {4},
  pages         = {044051},
  year          = {2016},
  doi           = {10.1103/PhysRevD.94.044051},
  eprint        = {1603.06587},
  archivePrefix = {arXiv}
}

@article{upadhye2012dark,
  author  = {Upadhye, A.},
  title   = {Dark energy fifth forces in torsion pendulum experiments},
  journal = {Physical Review D},
  volume  = {86},
  number  = {10},
  pages   = {102003},
  year    = {2012},
  doi     = {10.1103/PhysRevD.86.102003}
}

@article{yin2022experiments,
  author  = {Yin, P. and Li, R. and Yin, C. and Xu, X. and Bian, X. and Xie, H. and Duan, C.-K. and Huang, P. and He, J.-H. and Du, J.},
  title   = {Experiments with levitated force sensor challenge theories of dark energy},
  journal = {Nature Physics},
  volume  = {18},
  number  = {10},
  pages   = {1181--1185},
  year    = {2022},
  DOI     = {10.1038/s41567-022-01706-9} 
}

@article{almasi2015force,
  author  = {Almasi, A. and Brax, P. and Iannuzzi, D. and Sedmik, R. I. P.},
  title   = {Force sensor for chameleon and {Casimir} force experiments with parallel-plate configuration},
  journal = {Physical Review D},
  volume  = {91},
  number  = {10},
  pages   = {102002},
  year    = {2015},
  doi     = {10.1103/PhysRevD.91.102002}
}

@article{Zlatev:1999PRL,
  author        = {Zlatev, I. and Wang, L. and Steinhardt, P. J.},
  title         = {Quintessence, cosmic coincidence, and the cosmological constant},
  journal       = {Physical Review Letters},
  volume        = {82},
  number        = {5},
  pages         = {896--899},
  year          = {1999},
  doi           = {10.1103/PhysRevLett.82.896},
  eprint        = {astro-ph/9807002},
  archivePrefix = {arXiv}
}

@article{Copeland:2006IJMPD,
  author        = {Copeland, E. J. and Sami, M. and Tsujikawa, S.},
  title         = {Dynamics of dark energy},
  journal       = {International Journal of Modern Physics D},
  volume        = {15},
  number        = {11},
  pages         = {1753--1935},
  year          = {2006},
  doi           = {10.1142/S021827180600942X},
  eprint        = {hep-th/0603057},
  archivePrefix = {arXiv}
}

@article{Joyce:2015PhysRep,
  author        = {Joyce, A. and Jain, B. and Khoury, J. and Trodden, M.},
  title         = {Beyond the cosmological standard model},
  journal       = {Physics Reports},
  volume        = {568},
  pages         = {1--98},
  year          = {2015},
  doi           = {10.1016/j.physrep.2014.12.002},
  eprint        = {1407.0059},
  archivePrefix = {arXiv},
  primaryClass  = {astro-ph.CO}
}

@article{abdul2025desi,
  author        = {{DESI Collaboration} and Abdul Karim, M. and others},
  title         = {{DESI DR2} Results. {II}. {M}easurements of Baryon Acoustic Oscillations and Cosmological Constraints},
  journal       = {Physical Review D},
  volume        = {112},
  number        = {8},
  pages         = {083515},
  year          = {2025},
  doi           = {10.1103/tr6y-kpc6},
  eprint        = {2503.14738},
  archivePrefix = {arXiv},
  primaryClass  = {astro-ph.CO}
}

@article{burrage2018tests,
  author  = {Burrage, C. and Sakstein, J.},
  title   = {Tests of chameleon gravity},
  journal = {Living Reviews in Relativity},
  volume  = {21},
  number  = {1},
  pages   = {1},
  year    = {2018},
  doi     = {10.1007/s41114-018-0011-x}
}

@article{khoury2004chameleon,
  author  = {Khoury, J. and Weltman, A.},
  title   = {Chameleon cosmology},
  journal = {Physical Review D},
  volume  = {69},
  number  = {4},
  pages   = {044026},
  year    = {2004},
  doi     = {10.1103/PhysRevD.69.044026}
}

@article{burrage2015probing,
  author  = {Burrage, C. and Copeland, E. J. and Hinds, E. A.},
  title   = {Probing dark energy with atom interferometry},
  journal = {Journal of Cosmology and Astroparticle Physics},
  volume  = {2015},
  number  = {03},
  pages   = {042},
  year    = {2015},
  doi     = {10.1088/1475-7516/2015/03/042}
}

\newpage

\appendix
\renewcommand{\theequation}{A.\arabic{equation}}
\setcounter{equation}{0}

\section{Jacobi Elliptic Functions}
\label{sec : Apx - Jacobi Elliptic Functions}

The incomplete elliptic integrals of the first and second kind are respectively~\cite[Eq.~(17.2.6) and Eq.~(17.2.9)]{abramowitz1964handbook}\footnote{Elliptic integrals have multiple notation conventions, the simultaneous existence of which can lead to confusion. For example,~\cite{abramowitz1964handbook} mostly uses $F(u | m)$, which is not the same as $F(u, m)$, but is related as $F(u, m) \equiv F(u | m^2)$.}
\begin{align}
    \label{eq : Apx - Jacobi elliptic functions : Elliptic integral - incomplete first kind}
    F(\varphi, m) &:= \int_{0}^{\varphi} \frac{d \theta}{\sqrt{1 - m^2 \sin^2(\theta)}}, \\
    \label{eq : Apx - Jacobi elliptic functions : Elliptic integral - incomplete second kind}
    E(\varphi, m) &:= \int_{0}^{\varphi} \sqrt{1 - m^2 \sin^2(\theta)} \, d \theta.
\end{align}
The label `incomplete' means that the upper bound of the integral is itself an argument of the function, as opposed to the `complete' elliptic integrals where the value is fixed at $\pi/2$. For example, the complete elliptic integral of the first kind is defined as~\cite[Eq.~(17.3.2)]{abramowitz1964handbook}
\begin{equation}
    \label{eq : Apx - Jacobi elliptic functions : Elliptic integral - complete first kind}
    K(m) := F \left( \frac{\pi}{2}, m \right).
\end{equation}
The Jacobi elliptic functions are given as~\cite[Eq.~(16.1.5)]{abramowitz1964handbook}
\begin{align}
    \label{eq : Apx - Jacobi elliptic functions : sn (definition)}
    {\rm sn}(u, m) &:= \sin \left( F^{-1}(u, m) \right), \\
    \label{eq : Apx - Jacobi elliptic functions : cn (definition)}
    {\rm cn}(u, m) &:= \cos \left( F^{-1}(u, m) \right), \\
    \label{eq : Apx - Jacobi elliptic functions : dn (definition)}
    {\rm dn}(u, m) &:= \sqrt{1 - m^2 {\rm sn}^2(u, m)}.
\end{align}
From their trigonometric definitions we can relate the first two functions by
\begin{equation}
    \label{eq : Apx - Jacobi elliptic functions : identity - sn to cn}
    {\rm sn}^2(u, m) + {\rm cn}^2(u, m) \equiv 1.
\end{equation}
The derivatives of the Jacobi elliptic functions are~\cite[Eq.~(16.16.1-3)]{abramowitz1964handbook} 
\begin{align}
    \label{eq : Apx - Jacobi elliptic functions : sn derivative}
    \frac{d}{du}{\rm sn}(u, m) &= {\rm cn}(u, m) \, {\rm dn}(u, m), \\
    \label{eq : Apx - Jacobi elliptic functions : cn derivative}
    \frac{d}{du}{\rm cn}(u, m) &= - {\rm sn}(u, m) \, {\rm dn}(u, m), \\
    \label{eq : Apx - Jacobi elliptic functions : dn derivative}
    \frac{d}{du}{\rm dn}(u, m) &= - m^2{\rm sn}(u, m) \, {\rm cn}(u, m).
\end{align}
Combining these we get the derivative of a general product of Jacobi elliptic functions as
\begin{equation}
    \label{eq : Apx - Jacobi elliptic functions : derivative of general product}
    \frac{d}{du} \left[ {\rm sn}^a \, {\rm cn}^b \, {\rm dn}^c \right] = {\rm sn}^{a+1} \, {\rm cn}^{b+1} \, {\rm dn}^{c+1} \left[ \frac{a}{{\rm sn}^2} - \frac{b}{{\rm cn}^2} - m^2 \frac{c}{{\rm dn}^2} \right],
\end{equation}
where we have omitted the function arguments for reasons of space.

There exist other Jacobi elliptic functions, but these can be formulated both algebraically and symbolically in terms of the ones defined in Eqs.~(\ref{eq : Apx - Jacobi elliptic functions : sn (definition)}-\ref{eq : Apx - Jacobi elliptic functions : dn (definition)}). For ratios of different Jacobi elliptic functions, the naming rule used is
\begin{equation}
    \label{eq : Apx - Jacobi elliptic functions : ratio naming rule}
    {\rm pq}(u, m) := \frac{{\rm pn}(u, m)}{{\rm qn}(u, m)},
\end{equation}
where ${\rm p, q} \in \{ {\rm s, c, d} \}$~\cite[Eq.~(16.3)]{abramowitz1964handbook}. Note this means that the multiplicative inverse of any Jacobi elliptic function is simply the original name reversed. This extends to the original Jacobi elliptic functions as
\begin{equation}
    \label{eq : Apx - Jacobi elliptic functions : multiplicative inverse}
    {\rm ns}(u, m) := \frac{1}{{\rm sn}(u, m)}, \quad {\rm nc}(u, m) := \frac{1}{{\rm cn}(u, m)}, \quad {\rm nd}(u, m) := \frac{1}{{\rm dn}(u, m)},
\end{equation}
provided we treat `$\rm{nn}$' to be equivalent to unity.

The final naming convention we shall call upon is~\cite[Eqs.~(16.25)]{abramowitz1964handbook}
\begin{equation}
    \label{eq : Apx - Jacobi elliptic functions : integral squared naming rule}
    \begin{aligned}
    {\rm Pq}(u, m) &:= \int_{0}^{u} {\rm pq}^2(t, m) dt, \qquad \text{if q $\neq$ s}, \\
    {\rm Ps}(u, m) &:= \int_{0}^{u} \left[ {\rm ps}^2(t, m) - \frac{1}{t^2} \right] dt - \frac{1}{u}.
\end{aligned}
\end{equation}

Let us next consider how the Jacobi elliptic functions behave with imaginary arguments. If the first argument is imaginary we can use the relations~\cite[Eqs.~(16.20.1-3)]{abramowitz1964handbook}
\begin{align}
    \label{eq : Apx - Jacobi elliptic functions : argument u is imaginary (sn)}
    {\rm sn}(iu, m) &= i \, {\rm sc}\left( u, \sqrt{1 - m^2} \right), \\
    \label{eq : Apx - Jacobi elliptic functions : argument u is imaginary (cn)}
    {\rm cn}(iu, m) &= {\rm nc}\left( u, \sqrt{1 - m^2} \right), \\
    \label{eq : Apx - Jacobi elliptic functions : argument u is imaginary (dn)}
    {\rm dn}(iu, m) &= {\rm dc}\left( u, \sqrt{1 - m^2} \right),
\end{align}
while if the second argument is imaginary we have~\cite[Eqs.~(16.10)]{abramowitz1964handbook}
\begin{align}
    \label{eq : Apx - Jacobi elliptic functions : argument m is imaginary (sn)}
    {\rm sn}(u, i m) &= \nu \, {\rm sd}\left( \frac{u}{\nu}, \nu m \right), \\
    \label{eq : Apx - Jacobi elliptic functions : argument m is imaginary (cn)}
    {\rm cn}(u, i m) &= {\rm cd}\left( \frac{u}{\nu}, \nu m \right), \\
    \label{eq : Apx - Jacobi elliptic functions : argument m is imaginary (dn)}
    {\rm dn}(u, i m) &= {\rm nd}\left( \frac{u}{\nu}, \nu m \right),
\end{align}
where $\nu := 1/\sqrt{1 + m^2}$. The final identity involving imaginary arguments we will require is~\cite[Eq.~(17.4.8)]{abramowitz1964handbook}
\begin{equation}
    \label{eq : Apx - Jacobi elliptic functions : argument u is imaginary (F)}
    F \left( i \varphi, m \right) \equiv i F \left( \theta, \sqrt{1-m^2} \right), \qquad \text{if } \sinh\varphi = \tan\theta.
\end{equation}

\renewcommand{\theequation}{B.\arabic{equation}}
\setcounter{equation}{0}

\section{Correcting previous derivation}
\label{sec : Apx - Correcting previous derivation}

Part of the derivation outlined in Section~\ref{sec : Solution for N-layer system - Interior regions - Solution with a maximum} was based on the work in~\cite{ivanov2016exact}, which attempted to derive the chameleon field profile between two equally dense mirrors separated by an imperfect vacuum. The equivalent formula in this work is Eq.~(\ref{eq : interior region - maximum - field}). Since the mirrors have equal density, the field maximum must lie in the centre of the vacuum due to the symmetry of the system. Therefore, by recalling the definition of the $\zeta$-coordinate in Eq.~(\ref{eq : interior region - coorinate transformation (z to zeta)}) we have $\zeta_L = \zeta_R$ and so the field profile is
\begin{equation}
    \label{eq : appendix - correction - field profile}
    \phi = \phi_0 \, {\rm sn}^2 \left( {\rm Sn}^{-1} \left\{ \left[ {\rm Sn} \left( K(k), k \right) - \frac{L}{2 L_s} \left| 1 - \frac{2z}{L} \right| \right], k \right\}, k \right),
\end{equation}
where to be consistent with~\cite{ivanov2016exact} we have used $L_s := \sqrt{2} \phi_0^{3/2} / \Lambda^{5/2} = 2 \lambda_V k^{3/2}$, where $\lambda_V$ is the field's Compton wavelength in the vacuum, and translated our coordinate to position the origin at the left mirror boundary. However, comparing this to Eq.~(14) in~\cite{ivanov2016exact} we see that our results differ. By comparing our derivation with that in~\cite{ivanov2016exact}, both are equivalent up to and including Eq.~(\ref{eq : interior region - field gradient (maximum) - integrated}) (equivalently Eq.~(6) in~\cite{ivanov2016exact})\footnote{We acknowledge that there is an extra factor of $2$ in the corresponding expression in~\cite{ivanov2016exact}. This is due to a missing factor of $2$ that appears from the substitution of variables.}. In~\cite{ivanov2016exact} it is then argued that we can perform the replacement $k^2 \rightarrow p^2$, and then integrate over $p^2$. The problem with this approach is that we are defining a new equation, in an extended parameter space, for which we only have information when it coincides with our target problem at $p^2 = k^2$. To illustrate the issue this generates, consider the equation
\begin{equation}
    \label{eq : appendix - correction - modified integral express (p)}
        \int_{\arcsin(\sqrt{x})}^{\pi/2} \frac{\sin^2(\varphi) d \varphi}{\sqrt{1 - p^2 \sin^2(\varphi)}} = \frac{1}{L_s} \left( \frac{L}{2} - z \right) + \partial_{p^2} f \left(z, p^2 \right),
\end{equation}
where $f \left(z, p^2 \right)$ is an undetermined function. If we impose that $\partial_{p^2}f \left(z, p^2 \right)|_{ p^2=k^2 } = 0$ then Eq.~(6) in~\cite{ivanov2016exact} and Eq.~(\ref{eq : appendix - correction - modified integral express (p)}) above are consistent with one another when $p^2 = k^2$. This is then a valid extension of our target integral, which differs from~\cite{ivanov2016exact} if we choose $f \left(z, p^2 \right)$ to not be identically zero. Therefore, when we integrate over $p^2$, as is done in~\cite{ivanov2016exact}, we can get a non-zero contribution from $f \left(z, p^2 \right)$. Continuing in the same vein as~\cite{ivanov2016exact}, we arrive at the expression
\begin{equation}
    \label{eq : appendix - correction - integrated over p^2}
    E \left( \arcsin(\sqrt{x}), k \right) + \arccos(\sqrt{x}) = E \left( \frac{\pi}{2}, k \right) + \frac{k^2}{2 L_s} \left( \frac{L}{2} - z + L \, g(z) \right),
\end{equation}
where we have defined $g(z) := L_s \left[f \left(z, k^2 \right) - f \left(z, 0 \right)\right]/k^2L$ and used the incomplete elliptic function as defined in Eq.~(\ref{eq : Apx - Jacobi elliptic functions : Elliptic integral - incomplete second kind}). If we neglect $g(z)$ this is identical to Eq.~(10) in~\cite{ivanov2016exact}. In fact, if we made $z(x)$ the subject of Eq.~(\ref{eq : appendix - correction - integrated over p^2}), then $L \, g(x)$ would be the difference between the true position of the field value $x$ and that predicted by the solution in~\cite{ivanov2016exact}. If we had chosen to perform the sequence of manipulations in~\cite{ivanov2016exact}, but without replacing $k$ with $p$, we would then have to apply the Leibniz integral rule since one of the limits of integration, namely $\arcsin\left(\sqrt{x}\right)$, depends on $k$ through $x$. Therefore, we can equate the two approaches to conclude that $g(z)$ contains information regarding how $x$ depends on $k^2$ at a fixed value of $z$.

In principle we could derive the Taylor series expansion of $g(z)$. Evaluating Eq.~(\ref{eq : appendix - correction - integrated over p^2}) at $z=L/2$ we find $g(L/2)=0$. Taking the first derivative of Eq.~(\ref{eq : appendix - correction - integrated over p^2}) with respect to $z$, we find in the region $0 \leq z \leq L/2$ that
\begin{equation}
\label{eq : appendix - correction - g'(z)}
\begin{aligned}
    g'(z) &= \frac{2}{L} \left[ \frac{ \left(1 - \frac{k^2}{2} x \right) - \sqrt{1 - k^2 x} }{k^2 x} \right] \\
    &= \frac{1}{L} \sum_{n=1}^{\infty} \frac{(2n - 1)!!}{2^n (n+1)!} \left( k^2 x \right)^n,
\end{aligned}
\end{equation}
where on the second line we used the Taylor series expansion of $\sqrt{1 - k^2x}$. The Taylor series expansion of $g(z)$ can therefore be expressed as
\begin{equation}
\label{eq : appendix - correction - Taylor expansion of g(z)}
\begin{aligned}
    g(z) &= \frac{1}{L} \sum_{m=1}^{\infty} \frac{1}{m!} \left[ \partial_z^m g(z) \right]_{z=L/2} \left( z - \frac{L}{2} \right)^m \\
    &= \frac{1}{L} \sum_{m=1}^{\infty} \frac{1}{m!} \left[ \partial_z^{m-1} \left\{ \sum_{n=1}^{\infty} \frac{(2n - 1)!!}{2^n (n+1)!} \left( k^2 x \right)^n \right\} \right]_{z=L/2} \left( z - \frac{L}{2} \right)^m \\
    &= \frac{1}{L} \sum_{n=1}^{\infty} \frac{(2n - 1)!!}{2^n (n+1)!} k^{2n} \sum_{m=1}^{\infty} \frac{1}{m!} \left[ \partial_z^{m-1} \left\{ x^n \right\} \right]_{z=L/2} \left( z - \frac{L}{2} \right)^m \\
    &= - \sum_{n=1}^{\infty} \frac{(2n - 1)!!}{2^n (n+1)!} k^{2n} \left[ y_n \left( \frac{L}{2} \right) - y_n \left( z \right) \right],
\end{aligned}
\end{equation}
where on the first line we used that $g(L/2) = 0$, on the second we used Eq.~(\ref{eq : appendix - correction - g'(z)}) in place of $g'(z)$, and on the fourth line we introduced, through its Taylor series, the function
\begin{equation}
    \label{eq : appendix - correction - y_n(z) definition}
    y_n(z) := \frac{1}{L} \int_{0}^{z} x^n(t) \, dt.
\end{equation}
Since $x(z)$ is a monotonically increasing function with a value of unity at $z=L/2$ it must always be bounded by $0 < x^n(z) \leq 1$. Therefore, the function $y_n(z)$ must have bounds $(0, z/L)$, and so the square bracket on the last line of Eq.~(\ref{eq : appendix - correction - Taylor expansion of g(z)}) must obey
\begin{equation}
    \label{eq : appendix - correction - bounds on [y_n(L/2) y_n(z)]}
    0 \leq \left[ y_n \left( \frac{L}{2} \right) - y_n \left( z \right) \right] < \frac{1}{2}.
\end{equation}
Thanks to Eq.~(\ref{eq : appendix - correction - Taylor expansion of g(z)}) we then see that $g(z) \sim \mathcal{O}(k^2)$. Being more explicit, we can use the last line of Eq.~(\ref{eq : appendix - correction - Taylor expansion of g(z)}) with the identity
\begin{equation}
    \label{eq : appendix - correction - Taylor identity}
    \left( \frac{1 - \sqrt{1 - k^2}}{k} \right)^2 \equiv \sum_{n=1}^{\infty} \frac{(2n - 1)!!}{2^n (n+1)!} k^{2n},
\end{equation}
to rewrite the inequality in Eq.~(\ref{eq : appendix - correction - bounds on [y_n(L/2) y_n(z)]}) as
\begin{equation}
    \label{eq : appendix - correction - bounds on g(z)}
    0 \leq -g(z) < \frac{k^2}{2} \left( \frac{1 - \sqrt{1 - k^2}}{k^2} \right)^2 < \frac{k^2}{2}.
\end{equation}
So, for sufficiently small values of $k$, the deviation between the true solution and that derived in~\cite{ivanov2016exact} becomes negligible. Therefore, any existing results that were obtained using~\cite{ivanov2016exact} will not need correcting if $k \ll 1$. For example, prospective \textsc{Cannex} constraints were computed in~\cite{fischer2024screened}. Using Eqs.~(\ref{eq : interior region - maximum - constraint equation (k)}) and (\ref{eq : Example Solutions - three layers - field at walls (extrema)}) in tandem gives us a non-linear formula that can be solved for $k$. We find that much but not all of the parameter region constrained in~\cite{fischer2024screened} satisfies $k \ll 1$, and is therefore unaffected by the mistake in~\cite{ivanov2016exact}. In~\cite{ivanov2016exact} itself $k=0.070$, and~\cite{fischer2024numerical} has $\rho_V \rightarrow 0$ (therefore $k \rightarrow 0$), so both are unaffected.

As a final comment, we note that Eq.~(11) in~\cite{ivanov2016exact} was used incorrectly, as it applies to $E(u, m)$~\cite[Eq.~(17.2.10)]{abramowitz1964handbook}, but instead was used with $E(\varphi, m)$~\cite[Eq.~(17.2.9)]{abramowitz1964handbook}. Note these expressions evaluate to different values given the same arguments, but can be related~\cite[Eq.~(17.2.4)]{abramowitz1964handbook}.

\renewcommand{\theequation}{C.\arabic{equation}}
\setcounter{equation}{0}

\section{Matching to known approximate analytic solutions}
\label{sec : Apx - Matching to known approximate analytic solutions}

In~\cite{brax2011strongly}, an approximate analytic solution was derived for the chameleon field of a single plate next to a partial vacuum for the regime where $\phi \ll \phi^{\rm EV}_V$, where $\phi^{\rm EV}_V$ is the EV of the vacuum. We will now show how this approximate solution is consistent with the exact solution derived in Section~\ref{sec : Solution for N-layer system - exterior regions}.

To begin, recall that the last line of Eq.~(\ref{eq : Exterior regions - second integration}) gives the field profile in terms of $u := \sqrt{\phi/\phi^{\rm EV}_V}$ and $u_b := \sqrt{\phi_b/\phi^{\rm EV}_V}$, where $\phi_b$ is the field value on the boundary. Choosing our coordinates so the boundary is at $z = 0$ and the vacuum is located at $z > 0$, we write the field profile in the vacuum as
\begin{equation}
    \label{eq : Apx - Matching approximate solutions - single plate (exact)}
    -\frac{z}{2 \lambda_V} = \Delta + \frac{1}{2} \ln\left( 1 - \frac{\Delta}{1 - u_b} \right) - \frac{1}{2} \ln\left( 1 + \frac{\Delta}{1 + u_b} \right),
\end{equation}
where $\lambda_V$ is the field's Compton wavelength in the vacuum, and we have defined $\Delta := u - u_b$. Since the field is positive and monotonic, applying the approximation gives us the hierarchy $0 < u_b \leq u \ll 1$ near the plate, and hence $\Delta \ll 1 \pm u_b \approx 1$. We can therefore approximate Eq.~(\ref{eq : Apx - Matching approximate solutions - single plate (exact)}) as
\begin{equation}
\begin{aligned}
    -\frac{z}{2 \lambda_V}
    &= \Delta - \frac{1}{2} \sum_{n=1}^{\infty} \frac{(-\Delta)^{n}}{n} \left\{ \left( \frac{-1}{1 - u_b} \right)^n - \left( \frac{1}{1 + u_b} \right)^n \right\} \\
    &= \Delta - \frac{1}{2} \sum_{n=1}^{\infty} \frac{(-\Delta)^{n}}{n} \sum_{k=0}^{\infty} (-1)^k \binom{n+k-1}{k} \left[ (-1)^{n+k} - 1 \right] u_b^k \\
    &= \Delta - \sum_{n=1}^{\infty} \sum_{k=0}^{\infty} \delta^{(\rm odd)}_{n+k} \frac{1}{n} \binom{n+k-1}{k} \Delta^{n} u_b^k \\
    &\approx -\left[ \Delta u_b^2 + \Delta^2 u_b + \frac{1}{3} \Delta^3 \right] = \frac{1}{3} \left[u_b^3 - (u_b + \Delta)^3 \right] = \frac{1}{3} \left[ u_b^3 - u^3 \right],
\end{aligned}
\end{equation}
where in the first and second lines we used the Taylor series expansions for $\ln(1+x)$ and $(1+x)^{-n}$ respectively, and $\delta^{(\rm odd)}_{n+k}$ is unity if $n+k$ is an odd integer and zero otherwise. If we now note that the chameleon mass evaluated on the boundary is related to the $\lambda_V$ by $m_b = 1/(\lambda_V u_b^3)$, the above expression can be written as
\begin{equation}
    \label{eq : Apx - Matching approximate solutions - single plate (approx)}
    \phi \approx \phi_b \left( 1 + \frac{3}{2} m_b z \right)^{2/3}.
\end{equation}
Additionally, by taking the plate density to be $\rho_M \gg \rho_V$ and applying this to Eq.~(\ref{eq : junction conditions - harmonic mean field value}), we find that $\phi_b \approx 2 \phi_M$ (therefore $m_b \approx m_M/\sqrt{8}$). Both this formula for $\phi_b$ and Eq.~(\ref{eq : Apx - Matching approximate solutions - single plate (approx)}) agree with the result found in~\cite{ivanov2013influence, brax2011strongly} for $n=1$.

Let us now consider the case of a pair of parallel plates of density $\rho_L$ and $\rho_R$, separated by a vacuum of density $\rho_V \approx 0$. As $\rho_V \rightarrow 0$ the vacuum EV will diverge, but since we are in a bounded space the field maximum will tend to a finite value, $\phi_0$, meaning $k \rightarrow 0$. We therefore wish to apply $k \ll 1$ to Eq.~(\ref{eq : interior region - maximum - field}). To do this let us first note that for small $k$ we have $K(k) \approx \pi/2$~\cite[Eq.~(17.3.11)]{abramowitz1964handbook} and ${\rm sn}(u, k) \approx \sin(u)$~\cite[Eq.~(16.13.1)]{abramowitz1964handbook}. We therefore have
\begin{equation}
    {\rm Sn}(u ,k) \approx \int_0^u \sin^2(t) dt = \frac{1}{2} \left[ u - \sin(u) \cos(u) \right],
\end{equation}
\begin{equation}
    {\rm sn}^{-1}(u, k) \approx \sin^{-1}(u).
\end{equation}
Lastly, note that $\lambda_V k^{3/2} = \sqrt{\phi_0^3 / (2 \Lambda^5)} := \lambda_0$ is the Compton wavelength evaluated at $\phi_0$, and is independent of $\rho_V$. We can then write Eq.~(\ref{eq : interior region - maximum - field}) as
\begin{equation}
\label{eq : Apx - Matching approximate solutions - parallel plates (approx)}
    \begin{aligned}
        \frac{|z - z_0|}{\lambda_0}
        &= 2 \left[ {\rm Sn}\left\{ K(k), k \right\} - {\rm Sn}\left\{ {\rm sn}^{-1}\left(\sqrt{x}, k\right), k \right\} \right] \\
        &\approx \frac{\pi}{2} - \left[ \sin^{-1}\left(\sqrt{x}\right) - \sqrt{x} \cos\left(\sin^{-1}\left(\sqrt{x}\right)\right) \right] \\
        &\approx \cos^{-1}\left(\sqrt{x}\right) + \sqrt{x (1 - x)},
    \end{aligned}
\end{equation}
where we have used $z_0:= (\zeta_L - \zeta_R)/2$ and $x:=\phi/\phi_0$. We have therefore reproduced the approximate solutions found in~\cite{ivanov2013influence, burrage2016proposed} for a pair of parallel plates. Applying the same limit to Eq.~(\ref{eq : Example Solutions - three layers - field at walls (extrema)}) we find the vacuum boundary values are
\begin{equation}
    \phi_j = \phi^{\rm EV}_j \left[ 2 - \left( \frac{\phi^{\rm EV}_j}{\phi_0} \right) \right],
\end{equation}
where $j \in \{L, R\}$. If we now consider the scenario where the left plate is sufficiently dense, so that $\phi^{\rm EV}_L \ll \phi_0$, then $\phi_L \approx 2 \phi^{\rm EV}_L$, or equivalently $x_L \approx 2 \phi^{\rm EV}_L/\phi_0$. Next, we expand the right-hand side of the last line of Eq.~(\ref{eq : Apx - Matching approximate solutions - parallel plates (approx)}) as
\begin{equation}
    \cos^{-1}(\sqrt{x}) + \sqrt{x (1 - x)} = \frac{\pi}{2} - \frac{2}{3} x^{3/2} - \mathcal{O} \left( x^{5/2} \right).
\end{equation}
Therefore, if we evaluate Eq.~(\ref{eq : Apx - Matching approximate solutions - parallel plates (approx)}) at the left boundary, we find that an approximate expression for the location of the field's maximum is
\begin{equation}
\label{eq : Apx - Matching approximate solutions - z_0 (approx)}
    z_0 \approx - \frac{L}{2} + \frac{2}{3 m_L} \left[ \frac{3 \pi}{4} \left( \frac{\phi_0}{\phi^{\rm EV}_L} \right)^{3/2} - 2^{3/2} \right],
\end{equation}
where $m_L$ is the chameleon mass evaluated at $\phi^{\rm EV}_L$, $L$ is the width of the vacuum region (so by our choice of coordinate the location of the boundary with the left plate is $-L/2$). Note that this agrees with the equivalent result found in~\cite{burrage2016proposed} (see Eq.~(11) in~\cite{burrage2016proposed}) except for the sub-leading term. Using the numerical solver SELCIE~\cite{briddon2021selcie}, we verified our result against that in~\cite{burrage2016proposed}. After some investigation, we found that this disagreement comes from approximating the field inside the wall as quadratic. If the exact expression found in Eq.~(\ref{eq : Exterior regions - field gradient}) is used instead for the field gradient in the wall, we recover the sub-leading term found in Eq.~(\ref{eq : Apx - Matching approximate solutions - z_0 (approx)}).

Lastly, if we have $\rho_L = \rho_R$, then by the symmetry of the system we must have $z_0=0$. Applying this to Eq.~(\ref{eq : Apx - Matching approximate solutions - z_0 (approx)}) and neglecting the $2^{3/2}$ by recalling $\phi_0 \gg \phi^{\rm EV}_L$, we find that the field maximum inside the vacuum is
\begin{equation}
    \phi_0 \approx \Lambda \left( \frac{\sqrt{2} \Lambda L}{\pi} \right)^{2/3}.
\end{equation}
This agrees with the equivalent result in~\cite{ivanov2013influence} when evaluated with $n=1$.

\end{document}